\documentclass[fleqn,usenatbib]{mnras}
\usepackage[dvipsnames]{xcolor}
\usepackage{eso-pic}

\AddToShipoutPictureBG*{%
  \AtPageUpperLeft{%
    \hspace{0.75\paperwidth}%
    \raisebox{-1.\baselineskip}{%
      \makebox[0pt][l]{\textnormal{DES-2022-0708}}
}}}%

\AddToShipoutPictureBG*{%
  \AtPageUpperLeft{%
    \hspace{0.75\paperwidth}%
    \raisebox{-2.\baselineskip}{%
      \makebox[0pt][l]{\textnormal{FERMILAB-PUB-24-0104-PPD}}
}}}%

\usepackage{newtxtext,newtxmath}
\usepackage{mdframed}
\usepackage{graphicx}
\usepackage{subcaption}
\usepackage{lineno}
\usepackage{dsfont}
\let\oldequation\equation
\let\oldendequation\endequation
\renewenvironment{equation}
  {\linenomathNonumbers\oldequation}
  {\oldendequation\endlinenomath}

\usepackage{hyperref}
\usepackage[T1]{fontenc}
\usepackage{ae,aecompl}
\usepackage{todonotes}

\usepackage{graphicx}	
\usepackage{amsmath}	
\usepackage[dvipsnames]{xcolor}
\usepackage{tikz-cd}
\usepackage{mathtools}
\usepackage{amsfonts}
\usepackage{tabularx}

\definecolor{ork}{rgb}{0.9,0.1,0.3}
\definecolor{grbl}{rgb}{0.3,0.6,0.7}
\definecolor{bleu}{rgb}{0,0.5,0.6}

\newcommand{\Om}{\Omega_{\rm m}}

\newcommand{\healpix}[0]{\textsc{HEALPix}}
\newcommand{\pkdgrav}[0]{\textsc{pkdgrav3}}
\newcommand{\nbodykit}[0]{\textsc{nbodykit}}
\newcommand{\concept}[0]{\textsc{concept}}
\newcommand{\camb}[0]{\textsc{camb}}
\newcommand{\hyperrank}[0]{\textsc{hyperrank}}
\newcommand{\tarp}[0]{\textsc{tarp}}

\newcommand{\CLCNN}[0]{{$C_\ell \times$CNN}}
\newcommand{\CLPEAKS}[0]{{$C_\ell \times$Peaks}}

\newcommand{\mathd}{\ensuremath{\mathrm{d}}} 
\newcommand{\nside}[1]{$\textsc{nside} = #1$}
\newcommand{\pindex}[0]{\ensuremath{\phi}}

\title[DES Year 3: simulation-based $w$CDM inference]{Dark Energy Survey Year 3 results: likelihood-free, simulation-based $w$CDM inference with neural compression of weak-lensing map statistics}

\author[N. Jeffrey et al.]{
\parbox{\textwidth}{
\large{N.~Jeffrey,$^{1}$\thanks{E-mail: n.jeffrey@ucl.ac.uk}
L.~Whiteway,$^{1}$
M.~Gatti,$^{2}$
J.~Williamson,$^{1}$
J.~Alsing,$^{3}$
A.~Porredon,$^{4}$
J.~Prat,$^{5,6,7}$
C.~Doux,$^{2,8}$
B.~Jain,$^{2}$
C.~Chang,$^{6,7}$
T.-Y.~Cheng,$^{5}$
T.~Kacprzak,$^{9}$
P.~Lemos,$^{5}$
A.~Alarcon,$^{10,11}$
A.~Amon,$^{12,13}$
K.~Bechtol,$^{14}$
M.~R.~Becker,$^{10}$
G.~M.~Bernstein,$^{2}$
A.~Campos,$^{15}$
A.~Carnero~Rosell,$^{16,17,18}$
R.~Chen,$^{19}$
A.~Choi,$^{20}$
J.~DeRose,$^{21}$
A.~Drlica-Wagner,$^{6,22,7}$
K.~Eckert,$^{2}$
S.~Everett,$^{23}$
A.~Fert\'e,$^{24}$
D.~Gruen,$^{25}$
R.~A.~Gruendl,$^{26,27}$
K.~Herner,$^{22}$
M.~Jarvis,$^{2}$
J.~McCullough,$^{28}$
J.~Myles,$^{29}$
A. Navarro-Alsina,$^{30}$
S.~Pandey,$^{2}$
M.~Raveri,$^{31}$
R.~P.~Rollins,$^{32}$
E.~S.~Rykoff,$^{28,24}$
C.~S{\'a}nchez,$^{2}$
L.~F.~Secco,$^{7}$
I.~Sevilla-Noarbe,$^{33}$
E.~Sheldon,$^{34}$
T.~Shin,$^{35}$
M.~A.~Troxel,$^{19}$
I.~Tutusaus,$^{36}$
T.~N.~Varga,$^{37,38,39}$
B.~Yanny,$^{22}$
B.~Yin,$^{15}$
J.~Zuntz,$^{40}$
M.~Aguena,$^{17}$
S.~S.~Allam,$^{22}$
O.~Alves,$^{41}$
D.~Bacon,$^{42}$
S.~Bocquet,$^{25}$
D.~Brooks,$^{1}$
L.~N.~da Costa,$^{17}$
T.~M.~Davis,$^{43}$
J.~De~Vicente,$^{33}$
S.~Desai,$^{44}$
H.~T.~Diehl,$^{22}$
I.~Ferrero,$^{45}$
J.~Frieman,$^{22,7}$
J.~Garc\'ia-Bellido,$^{46}$
E.~Gaztanaga,$^{47,42,11}$
G.~Giannini,$^{48,7}$
G.~Gutierrez,$^{22}$
S.~R.~Hinton,$^{43}$
D.~L.~Hollowood,$^{49}$
K.~Honscheid,$^{50,51}$
D.~Huterer,$^{41}$
D.~J.~James,$^{52}$
O.~Lahav,$^{1}$
S.~Lee,$^{23}$
J.~L.~Marshall,$^{53}$
J. Mena-Fern{\'a}ndez,$^{54}$
R.~Miquel,$^{55,48}$
A.~Pieres,$^{17,56}$
A.~A.~Plazas~Malag\'on,$^{28,24}$
A.~Roodman,$^{28,24}$
M.~Sako,$^{2}$
E.~Sanchez,$^{33}$
D.~Sanchez Cid,$^{33}$
M.~Smith,$^{57}$
E.~Suchyta,$^{58}$
M.~E.~C.~Swanson,$^{26}$
G.~Tarle,$^{41}$
D.~L.~Tucker,$^{22}$
N.~Weaverdyck,$^{41,21}$
J.~Weller,$^{38,39}$
P.~Wiseman,$^{57}$
and M.~Yamamoto$^{19}$
\begin{center} (DES Collaboration) \end{center}
}
\parbox{\textwidth}{ \small
\textit{The authors' affiliations are shown in Appendix~\ref{append:affiliations}. \\
}}
}}
\date{Accepted XXX. Received 2024; in original form 2024}

\pubyear{2024}

\begin{document}
\label{firstpage}
\pagerange{\pageref{firstpage}--\pageref{lastpage}}

\maketitle

\begin{abstract}
We present simulation-based cosmological $w$CDM inference using Dark Energy Survey Year 3 weak-lensing maps, via neural data compression of weak-lensing map summary statistics: power spectra, peak counts, and direct map-level compression/inference with convolutional neural networks (CNN). Using simulation-based inference, also known as likelihood-free or implicit inference, we use forward-modelled mock data to estimate posterior probability distributions of unknown parameters. This approach allows all statistical assumptions and uncertainties to be propagated through the forward-modelled mock data; these include sky masks, non-Gaussian shape noise, shape measurement bias, source galaxy clustering, photometric redshift uncertainty, intrinsic galaxy alignments, non-Gaussian density fields, neutrinos, and non-linear summary statistics. We include a series of tests to validate our inference results. This paper also describes the \textit{Gower Street simulation suite}: 791 full-sky \pkdgrav{} dark matter simulations, with cosmological model parameters sampled with a mixed active-learning strategy, from which we construct over 3000 mock DES lensing data sets. For $w$CDM inference, for which we allow $-1<w<-\frac{1}{3}$, our most constraining result uses power spectra combined with map-level (CNN) inference. Using gravitational lensing data only, this map-level combination gives $\Om = 0.283^{+0.020}_{-0.027}$, ${S_8 = 0.804^{+0.025}_{-0.017}}$, and $w < -0.80$ (with a 68 per cent credible interval); compared to the power spectrum inference, this is more than a factor of two improvement in dark energy parameter ($\Omega_{\rm DE}, w$) precision.
\end{abstract}

\begin{keywords} gravitational lensing: weak -- cosmology: large-scale structure of Universe
\end{keywords}



\section{Introduction}

Weak gravitational lensing induces a pattern in the observed shapes of galaxies; we may use this to infer the distribution of foreground matter, including visible matter and (invisible) dark matter. The lensing effect is sensitive both to large scale structure formation and to geometric effects that probe the expansion history of the Universe.

Cosmological inference is typically performed using two-point correlation functions (e.g. power spectra) of the lensing signal. The currently most up-to-date analyses of this type are from the Dark Energy Survey (DES, \citealt{y3-cosmicshear1}; \citealt*{y3-cosmicshear2}), the Kilo-Degree Survey (KiDS, ~\citealt{Asgari_2021,li2023kids1000}), and Hyper Suprime-Cam (HSC, ~\citealt{hsc_shear}). Two-point statistics capture only some of the cosmologically relevant information and so are limited in discovery potential, in comparison to the information encoded in the full lensing \textit{mass map}; for DES Year 3 such lensing mass maps were presented in~\cite*{y3-massmapping}.

This paper has two scientific aims: (i) to use map-level inference to better constrain the cosmological parameters of the `$w$-Cold-Dark-Matter' ($w$CDM) model, and (ii) to use simulation-based inference (also known as  likelihood-free inference) methods to ensure realistic data modelling and reliable inference.

Deep learning methods (see \citealt{deep_learning} for an introduction) are used in two distinct ways in this analysis: \begin{enumerate}
\item Compression: we perform neural compression of high-dimensional data or summary statistics of the data; in our case we compress the map itself (using convolutional neural networks), the power spectra, and the peak counts from the map.
\item Neural likelihood estimation and validation: we use neural density estimation (as is typical with simulation-based inference) to learn the form of the likelihood from simulated mock data. We then validate the resulting posterior probability distributions.
\end{enumerate} 

This paper also serves as the public release of the \textit{Gower Street simulation suite}, consisting of 791 (so far -- the suite may grow in future) full-sky cosmological simulations that vary seven cosmological parameters of the $w$CDM model: the cosmological density parameter $\Omega_{\rm m}$, the amplitude parameter $\sigma_8$, the scalar spectral index $n_s$, the Hubble parameter $h=H_0/ (100 {\rm \ km \ s^{-1} \ Mpc^{-1}})$, the physical baryon density $\Omega_{\rm b} h^2$, the dark energy equation of state $w$, and the neutrino mass $m_{\nu}$ (the sum of the masses of the three neutrino mass eigenstates, quoted in electron volts). For the analysis in this paper, each full sky simulation can be split into four DES sky footprints, giving over 3000 quasi-independent mock DES surveys. Using multiple noise realizations we augment this suite to over $10^4$ non-independent mock DES surveys; these are used to train data compression and to perform simulation-based inference and posterior probability validation. 

One novel aspect of this work is the combination of simulation-based inference and map-level inference for an application with state-of-the-art weak gravitational lensing data. \cite{Fluri_2022} recently pioneered the use of deep learning for map-level weak lensing inference with KiDS data; this paper assumed a Gaussian likelihood. Other works have used machine learning methods to extract cosmological information, but without characterising the likelihood with simulation-based inference~\citep{Peel_2019, fluri_cnn, ribli_cnn, kids_cnn}. \cite{sv_lfi} used both simulation-based inference and deep learning for cosmological feature extraction, but this work used only the DES Science Verification data and so did not produce a competitive cosmological result. 

In a companion DES analysis, we are developing a simulation-based inference pipeline that uses wavelet scattering representations instead of convolutional neural networks, of which~\cite{gatti2023dark} is an initial description. In further analyses we will also try to understand the physical origin or environmental dependence of our map-level (deep learning) inference. These are all DES Year 3 analyses, awaiting the final full DES Year 6 data.

In section~\ref{sec:sbi} we introduce simulation-based inference, describing in turn the use of neural likelihood estimation to learn the form of the likelihood from realistic mock data (section~\ref{sec:nle}), the principle of data compression (section~\ref{sec:compression}), validation of the resulting posterior probability densities (section~\ref{sec:validation}), and parameter sampling and marginalization (section~\ref{sec:sampling}).

In section~\ref{sec:weaklensing} we give an overview of weak gravitational lensing and in section~\ref{sec:gowersims} we describe the Gower Street suite of simulations.

In section~\ref{sec:datamodel} we describe the DES Year 3 weak gravitational lensing data. We also describe how we generate mock DES data from the Gower Street simulations in a way that matches survey properties, noise, and forward modelling contributions to systematic uncertainty (e.g. intrinsic alignments of galaxies and photometric redshift uncertainty).

In section~\ref{sec:summarycompression} we describe each of the chosen summary statistics of the data and the data compression methods. The summary statistics described are the weak-lensing map itself (we describe how we construct convolutional neural networks to extract the cosmological information), the power spectra, and the counts of peaks in the lensing map. 

We present the cosmological inference results in section~\ref{sec:results} and conclude in section~\ref{sec:conclusion}.

\section{Simulation-based Inference}\label{sec:sbi}

\subsection{Motivation}

For parameter inference from complex physical systems, the likelihood, i.e. the conditional probability density $p(x | \theta)$ of the data $x$ given the model parameters $\theta$, is typically not known exactly or is too complex to be tractable. For these problems, \textit{simulation-based inference} (also known as `likelihood-free inference' or `implicit inference') provides a solution.

For weak gravitational lensing data, the exact form of the likelihood is typically not known. This is due both to the non-linear evolution of the cosmological density field and to several complicated observational effects (survey masks, various systematic biases, non-Gaussian noise contributions, etc.). Even if we assume that the underlying density field is Gaussian, the two-point statistics in weak lensing can have a significantly non-Gaussian distribution, especially if realistic observational effects are included \citep{alsing2017cosmological, sellentin_non_Gaussian, sellentin_skewed, taylor_delfi_lensing}.

For higher-order statistics, there is typically no closed-form expression for the likelihood. Even the expectation values of $p(x | \theta)$ for many higher-order statistics (e.g. peak counts) must be estimated from simulated mock data. We cannot expect the probability density $p(x | \theta)$ for these statistics to be Gaussian, as there are multiple sources of non-Gaussianity in the data model.

Even if the likelihood were known to be Gaussian, for observables that used simulated predictions (e.g. peak counts or map-level deep learning) the covariance matrix also has to be estimated from a significant number of simulations, typically run with fixed input parameters. The simulation-based inference approach avoids this, and hence can still be highly applicable even in the Gaussian likelihood case.

Furthermore, even if the likelihood is known, simulation-based inference methods allow implicit marginalization over nuisance parameters. As discussed in~\cite{momentnets}, traditional methods fail with large parameter spaces, whereas with simulation-based inference methods we can sidestep intractable high-dimensional inference and focus only on the selected parameters of interest. This implicit marginalization over nuisance parameters is central to the analyses presented in this paper, as we vary both unconstrained cosmological parameters and nuisance parameters (including $n(z)$ redshift distributions with $\sim 10^3$ dimensions). This is discussed further in section~\ref{sec:sampling}. 

\subsection{Neural likelihood estimation}\label{sec:nle}

This work uses the \textit{neural likelihood estimation} technique from the field of simulation-based inference; in this technique, the form of $p(x|\theta)$ is learned from mock data realizations~\citep{delfi1, MAF2}. By generating simulated mock data $x_i$, we are in fact drawing samples according to
\begin{equation}
    x_i \sim p(x | \theta_i)
\end{equation}
\noindent where $\theta_i$ are the input parameters to the simulation with index $i$.
From a set of simulated mock data labelled by their parameter values $\{x_i, \theta_i\}$, we can then learn a density $q$ that approximates the underlying probability density $p$, such that $p(x | \theta) \approx q(x | \theta)$.

In our case, $\theta$ is a chosen subset of the $w$CDM model parameters coupled with nuisance parameters corresponding to observational effects (e.g. intrinsic alignment amplitude).

Given parameters of interest $\theta$ and given some data $x$ (e.g. the lensing map or its power spectrum), our first step is to estimate $p(x | \theta)$. This estimated likelihood is then evaluated for the observed data $x_O$, from which as usual the posterior probability density of the parameters can be related to the likelihood via Bayes' theorem:
\begin{equation}
p( \theta | x_O ) = \frac{p(x_O | \theta) \ p(\theta)}{p(x_O)} \ \ .
\end{equation}

To estimate the conditional distribution $p( x | \theta)$, we use the pyDELFI~\citep{delfi2} package\footnote{\url{https://github.com/justinalsing/pydelfi}} with an ensemble of neural density estimators (NDEs). NDEs use neural networks to parameterize densities, including (as here) \textit{conditional} probability densities.

An NDE gives an estimate $q  (x |  \theta, {\varphi})$ by varying the ${\varphi}$ neural network parameters (e.g. weights and biases) to minimize the loss function

\begin{equation} \label{eq:delfi_loss}
U({\varphi}) = - \sum_{i=1}^N \log q  ({x}_i |  {\theta}_i ; {\varphi}) \ \ 
\end{equation}

\noindent over the $N$ forward-modelled mock data $x_i$. This loss corresponds to minimizing the Kullback-Leibler divergence~\citep{kullback1951}, a measure of change from the estimate $q$ to the target $p$.

We have available two types of NDEs: Gaussian Mixture Density Networks (MDN;~\citealt{mdn}) and Masked Autoregressive Flows (MAF;~\citealt{MAF}). An MDN represents the conditional density as a sum of several Gaussian components. A MAF is a type of Normalizing Flow i.e. it uses a series of bijective transformations from simple known densities (e.g. standard Gaussian) to the target density \citep{normalizing_flows,kingma2016improved,MAF2}.

For further details see \cite{sv_lfi} (in which a similar neural likelihood estimation setup was used, and in which may be found a more technical introduction to these NDE methods).

However, unlike \cite{sv_lfi}, the results presented in this paper use only MAFs, as these were found to perform better at hard prior boundary edges (e.g. for $w \approx -1$). The MDNs were used only for validation with simulated data analyses. For the presented results (section~\ref{sec:results}) we use an ensemble of four MAFs: each had either three, five, or six transformations (Masked Autoencoder for Distribution Estimation, i.e. MADE) with each using a neural network with two hidden layers (with widths of either 40 or 50).

\subsection{Principle of data compression}\label{sec:compression}

Density estimation of $p( x | \theta)$ rapidly increases in difficulty as the dimensionality $\mathrm{dim}(x)$ of the data vector $x$ increases (the `curse of dimensionality'). In this DES weak-lensing analysis, the data dimensionality is $\sim 10^7$ for the case of map-level inference and $\sim 10^3$ for inference using power spectra and peak counts. Direct estimation of $p( x | \theta)$ is intractable.

We take the (now standard) approach of data compression: apply some function $\mathcal{F}$ to the data to return compressed data $t = \mathcal{F}(x)$, while trying to preserve information about the parameters $\theta$. 

A poor compression (i.e. one that loses information) will not lead to biased inference. Because the same compression is applied consistently to both the simulated data and the observed data, a less-informative summary statistic $t_{\rm lossy}$ will lead to inflated posterior distributions on $\theta$. In the limit of uninformative compression, any posterior distribution $p(\theta | t_{\rm lossy})$ will merely be equal to the prior $p(\theta)$.

Although we do not have to worry about poor compression leading to incorrect inference, we clearly want to find a compression scheme that is maximally informative with respect to the parameters of interest $\theta$.  Different techniques are available for compression, all of which aim to maximize the information content of $t$ while dramatically reducing the dimensionality.

Under certain conditions it is possible to find $\mathcal{F}$ for which the dimension of $t$ equals the number of inferred parameters,  $\mathrm{dim}(t)=\mathrm{dim}(\theta)$, and which also is lossless with respect to the Fisher information~\citep[e.g.][]{moped, alsing_compression}. 

Neural compression, which we use in this DES analysis, takes advantage of the flexibility of neural networks to parameterize $\mathcal{F}$. The neural network is trained using simulated mock data. Existing methods include the Information Maximizing Neural Network~\cite[IMNN;][]{imnn}, which maximizes the Fisher information, and Variational Mutual Information Maximization~\cite[VMIM;][]{sv_lfi}, which maximizes the mutual information between the compressed data and the target parameters.

Instead of these methods we use a mean-square error (MSE) loss function to compress the data. This corresponds to an estimate of the mean of the posterior distribution for each parameter. Such a point estimate is clearly informative about the target parameters, and can be contrasted with the maximum likelihood parameter estimate, which corresponds to an optimal score compression \citep[with some caveats:][]{alsing_compression}. We do not expect this MSE compression to be optimal (e.g. compared to VMIM), but it is simple to implement.

The network architecture for the compression used in this work is described in detail in section~\ref{sec:summarycompression}. As the MSE only depends on the marginal posterior per parameter, we train a different network per parameter. Multiple noise realizations serve as data augmentation in our training data for compression. Throughout this analysis, the neural compression is learned from different noise realizations of the mock data to those that are used for neural likelihood estimation -- this is to avoid over-fitting. 

\subsection{Posterior probability validation}\label{sec:validation}

\subsubsection{Coverage tests} \label{sec:coverage_test_theory}

Coverage tests in Bayesian analysis check whether credible intervals have the expected probabilities. Looking at one-dimensional marginalized posteriors, we define a particular \textit{credible interval} to be the narrowest interval containing (say) $90\%$ of the probability weight; other credible intervals would work equally well. (This can be generalized e.g. \citealt{lemos2023sampling}). View the inference process as a procedure which, given observed data $x_O$, yields a posterior distribution $p(\theta|x_O)$ and hence a credible interval for $\theta$. In the coverage test we use a parameter $\theta_{\textrm{test}}$, selected from the prior $p(\theta)$, as input to a simulation yielding output data $x_{\textrm{test}}$, from which we derive a posterior $p(\theta|x_{\textrm{test}})$ and hence a credible interval; if the inference process is correctly implemented then the true test parameter value $\theta_{\textrm{test}}$ will fall in this credible interval $90\%$ of the time. By repeating with many such $\theta_{\textrm{test}}$ we are able to gain confidence that our estimated posterior distributions are indeed correct~\citep[][]{prangle2013diagnostic, sbi_crisis}.

This test is relatively straightforward for this type of simulation-based inference, for which we have a number of existing mock data simulations and where the inference scheme is \textit{amortized} (and so fast to evaluate probabilities for new data). Coverage testing is a useful aspect of inference, ensuring that the results are reliable, which is often unfeasible with traditional statistical approaches.

Given the computational expense of each simulation giving us a limited supply of mock data realisations, the biggest risk of failure is that we have insufficient simulations to robustly estimate the likelihood. Coverage tests can reassure us that we have sufficient numbers of simulations for this task; a successful coverage test implies there were enough simulations to accurately estimate the likelihood.

In this analysis we show successful coverage tests for inference using our learned likelihoods; this serves as one validation of the posterior distribution obtained for the actual observed data $p(\theta | x_O)$.

\subsubsection{Neural density ensemble convergence} \label{sec:ndeconvergence}

The individual likelihood estimates from the neural density ensemble can be used as a further validation step. The individual density estimates will converge to a common value as the number of simulations increases; therefore, if the posterior distributions from each independent density estimation are in disagreement, this would be evidence that we had an insufficient number of simulated mock data realizations.  

\subsection{Parameter sampling \& marginalization}\label{sec:sampling} 

The main strength of neural \textit{likelihood} estimation (learning $p(x |\theta)$) rather than the neural \textit{posterior} estimation (learning $p(\theta | x)$) is that the parameters $\theta$ in the training data (the simulations) do not have to be drawn from the prior $p(\theta)$.

This has two benefits. The first is that the prior can be changed at will after the simulations have been run (for example, to take new external information into account). The second, of particular importance to this work, is that additional simulations can be run in regions of parameter space that are most useful for the neural density estimation; this is known as \textit{active learning}. One can choose the parameter values for the new simulation from some \textit{acquisition function}, which may be based on the existing posterior estimates, to improve robustness. In this DES analysis, this was implemented in two stages: (1) most $\sigma_8$ and $\Om$ parameters were at first distributed according to the existing DES analysis constraints, and (2) after an initial simple blind power spectrum analysis, new simulations were run with $\sigma_8$ and $\Om$ values (known only to the computer) in regions of parameter space with high NDE ensemble variance (see section~\ref{sec:ndeconvergence}). Our sampling scheme is discussed further in section~\ref{sec:gowersims}.

For the parameters that are not part of the active learning scheme, we can still choose to distribute them according to a prior. If any set of parameters $\theta_{\rm marginal}$ is distributed according to the chosen prior $p(\theta_{\rm marginal})$, and if these parameters are excluded from the parameter set $\theta$ used for neural likelihood estimation, then these $\theta_{\rm marginal}$ parameters will be implicitly marginalized during inference. This is explained via \textit{marginal posterior density} estimation in ~\cite{momentnets}. The uncertainty in these parameters $\theta_{\rm marginal}$ is still accounted for in the resulting posterior distributions, as the parameters are varied in each simulated mock data realization, but the parameters are implicitly marginalized, which avoids explicit (and intractable) high-dimensional density estimation and unnecessary marginalization integration.

In the Gower Street simulations, all parameters other than $\sigma_8$ and $\Om$ are drawn from their prior distributions (with some caveats; section~\ref{sec:gowersims}). All observational nuisance parameters (e.g. intrinsic alignment and redshift) are also drawn from their priors in the mock DES lensing map generation (section~\ref{sec:datamodel}). This allows implicit marginalization if necessary. 

\section{Weak gravitational lensing} \label{sec:weaklensing}

Weak gravitational lensing (WL) \citep{Bartelmann2001} is the coherent slight alteration (for us, primarily \textit{shearing}) of the shapes of distant `source' galaxies by the gravitational influence of intervening matter (mostly dark). The unlensed shapes are unknown, and thus act as a noise term in WL analysis; the large surface density of galaxies visible in modern surveys allows the WL convergence signal (a weighted average of the overdensity along the line-of-sight, convolved with the redshift density of source galaxies) to be measured despite this noise. The two-point correlation functions of the signal may be estimated from observed data and compared to theoretical predictions from a cosmological model, thereby constraining the parameters of the model. Alternatively, convergence \textit{maps} may be constructed. The convergence field is not a Gaussian random field, and so the power spectrum is not the full story; there is further information in various beyond-two-point statistics from these maps. Alas, theoretical model predictions for these so-called \textit{non-Gaussian} statistics are generally not available; results derived from simulations must be used instead.

This section describes briefly the theory required to link the WL shear observables to results obtained from theory or simulations.

\subsection{Theory} \label{ssec:weaklensingtheory}

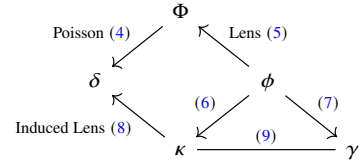
\begin{figure}
\centering
\begin{tikzcd}
    & \Phi\arrow{dl}[swap]{\textrm{Poisson }(\ref{eq:poisson})}  \\
    \delta & & \phi\arrow{ul}[swap]{\textrm{Lens }(\ref{eq:lens})}\arrow{dr}{(\ref{eq:gammaphi})}\arrow{dl}[swap]{(\ref{eq:kappaphi})} \\
    & \kappa\arrow{ul}{\textrm{Induced Lens }(\ref{eq:kappadelta})}\arrow[dash]{rr}{(\ref{eq:kappagamma})} & & \gamma
\end{tikzcd}
\caption{The relationships among weak-lensing fields $\Phi$ (gravitational potential), $\delta$ (overdensity), $\phi$ (weak-lensing potential), $\kappa$ (convergence), and $\gamma$ (shear); these relationships allow us to link observations to cosmological theory and simulations. Arrows represent spatial second derivatives; the line between $\kappa$ and $\gamma$ is a relationship of harmonic coefficients. Numbers refer to the corresponding equations in section~\ref{ssec:weaklensingtheory}, where further details are given.}
\label{fig:weaklensingtheory}
\end{figure}

We follow \cite*{y3-massmapping} section 2; see that paper for full details. See Figure \ref{fig:weaklensingtheory} for a schematic diagram of the relationships between the fields discussed.

The gravitational potential $\Phi$ and the matter overdensity field $\delta \equiv \rho/\bar{\rho}-1$ are related by the Poisson equation
\begin{equation}
\label{eq:poisson}
\nabla^2_r \Phi(t, \boldsymbol{r}) = \frac{3 \Om H_0^2}{2 a(t)} \delta(t, \boldsymbol{r}) \ .
\end{equation}
Here $\boldsymbol{r}$ is a comoving spatial coordinate and $a$ is the scale factor.

The weak-lensing potential $\phi$ is defined via the lens equation; $\phi$ is sourced by the gravitational potential, together with a lensing efficiency factor (written here assuming a flat Universe), all integrated along the line of sight to a source galaxy at comoving distance $\chi$ (here we use the Born approximation), and then further integrated over the redshift distribution $n(z)$ of source galaxies:

\begin{equation}
\label{eq:lens}
\phi (\theta,\varphi) = \frac{2}{c^2} \int_0^{\infty} \mathd \chi \ n(z(\chi)) \ \int_0^{\chi} \mathd \chi' \frac{(\chi -\chi')}{\chi \chi'} \Phi (\chi',\theta,\varphi) .
\end{equation}

The weak lensing potential is defined on the celestial sphere, so it is convenient to use the formalism of spin-weight functions on the sphere; see \cite{castro2005weak} for details and see also \cite{sellentin2023almanac} Appendix A for geometrical comments. Let $\eth$ and $\bar{\eth}$ denote the spin-weight covariant derivative and its adjoint. Let $\kappa$ and $\gamma$ be the weak-lensing convergence (spin-weight 0) and shear (spin-weight 2); they are second derivatives of the weak-lensing potential:

\noindent\begin{tabularx}{\columnwidth}{@{}XX@{}}
\begin{equation}
\kappa = \frac{1}{4}(\eth\bar{\eth}+\bar{\eth}\eth)\phi
\label{eq:kappaphi}
\end{equation}
&
\begin{equation}
\textrm{and } \quad \gamma = \frac{1}{2}\eth\eth \phi.
\label{eq:gammaphi}
\end{equation}
\end{tabularx}

\noindent Eqs \ref{eq:poisson}, \ref{eq:lens}, and \ref{eq:kappaphi} yield an induced lens equation linking $\delta$ and $\kappa$:
\begin{equation} 
\kappa(\theta, \phi) = \frac{3 \Om H_0^2}{2 c^2} \int_0^{\infty} \mathd \chi n(z(\chi)) \int_0^{\chi} \mathd \chi' \frac{\chi' (\chi -\chi')}{\chi} \frac{\delta(\chi', \theta, \phi)}{a(\chi')}.
\label{eq:kappadelta}
\end{equation}

Finally we move to harmonic space, representing an arbitrary field $b$ by its coefficients $b_{\ell m}$ with respect to the basis of spherical harmonic functions of the appropriate spin-weight. Now $\eth$ and $\bar{\eth}$ behave in a simple fashion in this basis, and so Eqs \ref{eq:kappaphi} and \ref{eq:gammaphi} yield
\begin{equation}
\label{eq:kappagamma}
\gamma_{\ell m} = -\sqrt{\frac{(\ell-1)(\ell+2)}{\ell(\ell+1)}} \kappa_{\ell m}.
\end{equation}

Using $\kappa$ as a link, Eqs \ref{eq:kappadelta} and \ref{eq:kappagamma} together connect $\gamma$ (which may be measured using the shapes of source galaxies -- taking into account shape noise and other sources of noise, partial sky coverage, and systematic effects such as intrinsic alignments) with $\delta$ (which may be treated theoretically -- at least up to two-point statistics -- or modelled via simulations). We thus have the desired link between theory (or simulations) and observations.

\begin{figure*}
  \includegraphics[width=0.8\textwidth]{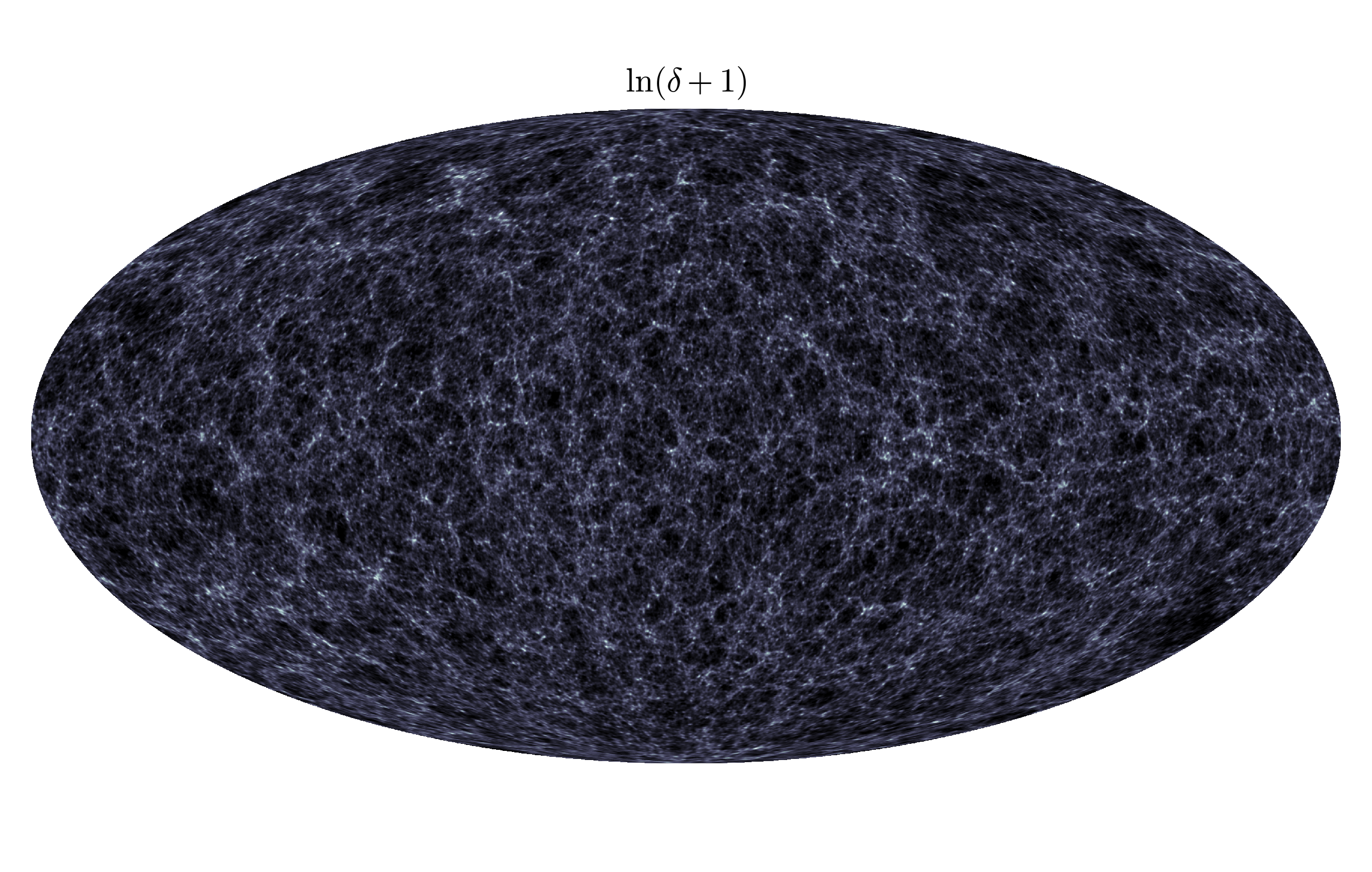}
 \vspace{-0.8cm}
 \caption{Example dark matter simulation from the Gower Street simulation suite. This map on the celestial sphere (Mollweide projection) uses the average overdensity $\delta$ from all shells up to a redshift $z=0.15$. Such simulations form the basis of the mock DES Y3 weak lensing maps used in the inference pipeline. \label{fig:gower_street}}
\end{figure*}

\begin{figure}
\includegraphics[width=0.9\columnwidth]{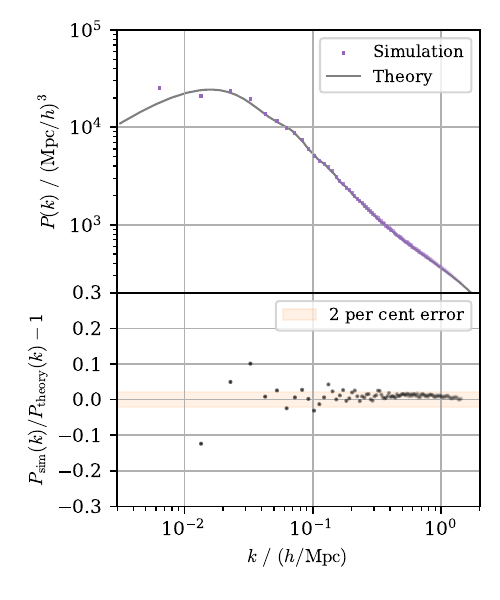}
\caption{Comparing the matter power spectrum $P(k, z=0)$ of a simulation to that from theory. The theory prediction for the power spectrum $P_{\rm theory}(k)$ combines \camb{} for linear theory \citep{camb} and the Euclid emulator~\citep{euclidemu} for the non-linear contribution. At small scales, where the finite resolution of the simulation is expected to cause inaccuracies, the systematic error remains below 2 per cent. \label{fig:matterpower}}
\end{figure}

\section{Gower Street simulations} \label{sec:gowersims}

\subsection{Simulation configuration}

The Gower Street suite of simulations consists of 791 gravity-only full-sky $N$-body simulations, produced using the \pkdgrav{} code \citep{potter2017pkdgrav3}, spanning a seven-dimensional parameter space in $w$CDM ($\Omega_{\rm m}$, $\sigma_8$, $n_s$, $h$, $\Omega_{\rm b}h^2$, $w$, $m_{\nu}$). 

For reviews of the theory of simulations, see \cite{Efstathiou1985} and \cite{Angulo2022}. In common with other $N$-body simulation codes, \pkdgrav{} uses a box of side $L$, filled with $N^3$ particles. At a start time, corresponding to redshift $z_0$, the particles are arranged in phase space (positions and velocities) so as to match desired initial conditions; for this, \pkdgrav{} uses second-order Lagrangian Perturbation Theory (2LPT). The positions/velocities of the particles are then updated (under the influence of gravity -- modelled as Newtonian -- against the backdrop of Universe expanding according to specified cosmological parameters) to yield snapshots of positions/velocities at various discrete times.

From this four-dimensional dataset, \pkdgrav{} extracts \textit{lightcone} data i.e. it restricts the dataset to events currently visible to the theoretical observer at the centre of the simulation. Specifically, the code estimates each particle's worldline (by interpolating between the particle's known positions at each time slice) and calculates, in four-dimensional space, the intersection of this worldline (a one-dimensional curve) with the observer's lightcone (a three-dimensional cone). This gives, for each particle, what event (redshift and position) on its worldline is currently visible. These data are then binned, by redshift (a bin corresponds to the redshift interval between two snapshots) and by position on the sky (into \healpix{} \citep{gorski2005} pixels). The results (particle count per pixel per redshift bin) are then output, with one file per redshift bin. For higher redshifts the comoving distance to the redshift will exceed the box side $L$; to avoid this, the simulation box is replicated $M$ times in each direction (a total of $(2M)^3$ replications, with the observer at the centre of this `super-box').

The Gower Street simulations use $L = 1250 \, h^{ - 1}$ Mpc and $N = 1080$. We set the initial redshift to $z_0 = 49$, and we produce 101 snapshots (and hence 100 lightcone files), equally spaced in proper time between $z_0$ and redshift zero. For the \healpix{} pixelization we set \nside{2048}. The simulation box is replicated $M=10$ times\footnote{We thank Janis Fluri for code amendments allowing an increase from the default $M=3$ value.}, although the bulk of our redshift distributions ($z<1.5$) can be covered by only three replications.

Fig.~\ref{fig:gower_street} presents an example of such a simulation; the map shows the matter overdensity as derived from the \pkdgrav{} particle count.

To validate the output of our simulations, we saved the three-dimensional particle positions as a final redshift snapshot at $z=0$ for a single simulation run whose parameters were $\Omega_{\rm m} = 0.3001$, $\sigma_8 = 0.7894$, $n_s = 0.95$, $h = 0.687$, $\Omega_{\rm b}h^2 = 0.02243$, $w = -0.95$, $m_{\nu} = 0.065$. Limits on computation time and disk space prevented us from generating these for multiple simulations. From this snapshot we measured the matter power spectrum using the \nbodykit{} ~\citep{hand2018nbodykit} code. Fig.~\ref{fig:matterpower} compares a) the measured power spectrum from this simulation to b) the theoretical power spectrum calculated using the Euclid emulator~\citep{euclidemu} code. At small scales, where the finite resolution of the simulation would be expected to cause inaccuracies, the difference between the measured and theoretical power spectrum remains below 2 per cent. This is within the relative error between different non-linear power spectrum prescriptions and other modelling choices, such as choice of neutrino model or astrophysical feedback model. Baryon feedback effects are not included in the simulation suite, but their effects are tested in section~\ref{sec:results}.

This paper serves as the formal release of these simulations, which are available at \url{www.star.ucl.ac.uk/GowerStreetSims/}.

\subsection{Cosmological parameters}
\label{ssec:cosmoparams}

A total of $791$ simulations were performed. The first $192$ of these were `verification' runs, done to test the software pipeline; they had a naive handling of neutrinos. For these runs, the initial conditions were specified to \pkdgrav{} via a transfer function generated using \nbodykit{} \citep{hand2018nbodykit}; in all of these runs the neutrino mass was fixed to $0.06$. Neutrinos played no role beyond this in these initial simulations. Further simulations, beyond these initial verification runs, had a more sophisticated handling of neutrinos, done via the \concept{} software \citep[see][]{tram2019}.

Each simulation was given its own values for seven cosmological parameters within $w$CDM: $\Omega_{\rm m}$, $\sigma_8$, $n_s$, $h$, $\Omega_{\rm b}h^2$, $w$, and $m_{\nu}$; in addition, each simulation had a different value for the random seed used when generating initial conditions (so that the simulations also display a range of behaviours arising from cosmic variance). The two parameters for which weak-lensing observations are most constraining, $\Omega_{\rm m}$ and $\sigma_8$, were chosen via \textit{active learning}: during later runs, values for these parameters were chosen so to maximize the incremental constraining power of the simulation suite (i.e. concentrated in regions of parameter space that were both important and under-represented). For simplicity, this was done simply by sampling these parameters from the posterior distribution of the parameters (both from the existing published DES results and as calculated using the simulations so far -- see \citealt{alsing2018optimal}), but with a hard exclusion zone around already-used parameter combinations.

The remaining parameters were chosen (independently) as follows:
\begin{itemize}
\item{$n_s \sim \mathcal{N}(0.9649, 0.0063)$; from Planck \citep{aghanim2020planck} but with the standard deviation boosted by a factor of 1.5.} 
\item{$h \sim \mathcal{N}(0.7022, 0.0245)$; consistent with SH0ES~\citep{Riess_2022} and Planck \citep{aghanim2020planck}, as its mean is midway between the means of these experiments, and its one standard deviation contour encompasses the two standard deviation contours of both experiments.}
\item{$\Omega_{\rm b}h^2 \sim \mathcal{N}(0.02237, 0.00015)$; from Planck \citep{aghanim2020planck}.}
\item{$w \sim \mathcal{N}(-1, 1/3)$, but with values less than $-1$ or greater than $-1/3$ then discarded. However, for the first 128 runs (part of the `science verification' runs), this discarding was not done (resulting in approximately 64 runs with $w<-1$). These runs have been kept as they help to smooth what would otherwise be the discontinuity at $w=-1$.

This choice of $w>-1$ excludes phantom dark energy. This has some theoretical justification for an N-body simulation on an expanding background, but is also motivated by computational limitations with low values of $\Om$ when using \pkdgrav{} and \concept{}.}
\item{$m_{\nu}$: As described in more detail above, fixed at $0.06$ for the initial 192 simulations and with $\log(m_{\nu}) \sim \mathcal{U}[\log(0.06), \log(0.14)]$ thereafter.}
\end{itemize}
In the above, $\mathcal{N}(\mu,\sigma)$ denotes a normal distribution with the indicated mean and standard deviation and $\mathcal{U}[a, b]$ denotes a uniform distribution with the indicated limits.

The parameter values are not sampled through i.i.d. draws from their respective distributions. Instead, to avoid similar parameter combinations arising due to random chance, we sample a multi-variate uniform distribution using a mixture of Sobol and Halton sequences (and then transform where necessary from uniform to Gaussian by applying the inverse function of the Gaussian cumulative distribution function).  

\begin{figure*}
  \includegraphics[width=0.8\textwidth]{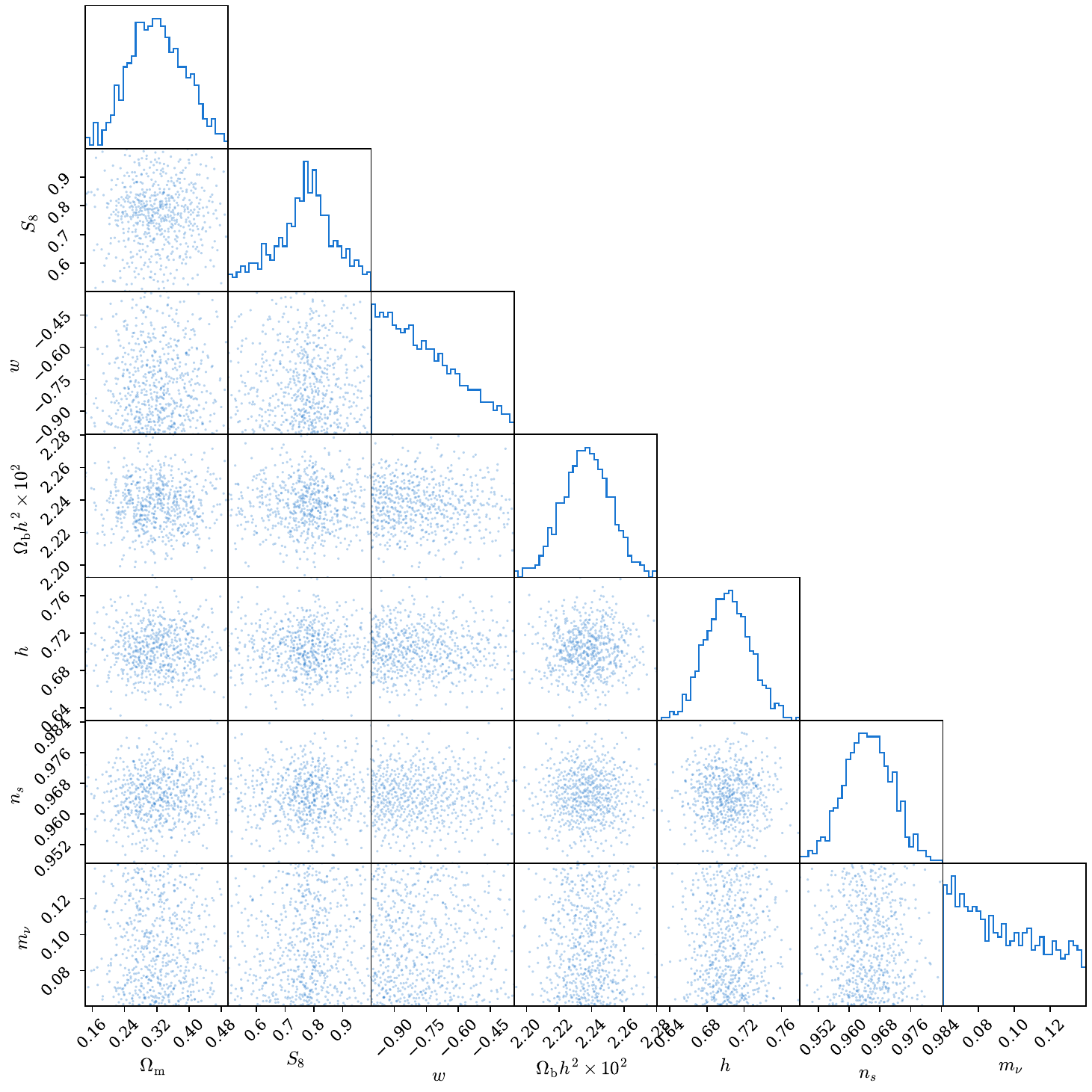}
 \caption{Parameter values used in the Gower Street simulations. The cosmological parameters span variations the in $\nu w$CDM model. We exclude the initial 192 'verification' runs to simplify the presentation of the neutrino mass distribution. For all parameters other than $\Om$ and $S_8$, these parameters are distributed according to their prior probability distributions. For $\Om$ and $S_8$, we always use neural likelihood estimation to condition on these parameters, removing the dependence on their simulated distribution.}
 \label{fig:gower_street_params}
\end{figure*}

Fig.~\ref{fig:gower_street_params} displays the parameter values for pairs of parameters used in the Gower Street simulations. We only show the parameter combinations for the simulations not included in the initial 192 'verification' runs, so that we can include the neutrino mass in the figure.

\section{Dark Energy Survey data and mock data modelling} \label{sec:datamodel}

\subsection{DES Year 3 weak lensing data}

DES is a photometric galaxy survey that covers $\sim 5000~\mathrm{\deg}^2$ of the South Galactic cap. Mounted on the Cerro Tololo Inter-American Observatory four metre Blanco telescope in Chile, the $570$~megapixel Dark Energy Camera \citep[DECam,][]{decam} images the field in $grizY$ filters. We use data from the first three years of the survey (DES Y3).

The simulated galaxy catalogues are created so as to match DES Y3 for \textit{known} properties. For example, the sky mask is known but the intrinsic alignment model parameters are not, so we simulate with fixed sky mask but vary the intrinsic alignment amplitude in each simulation.

The DES Y3 shear catalogue \citep*{y3-shapecatalog}, built upon the Y3 Gold catalogue \citep{y3-gold}, uses the \texttt{METACALIBRATION} algorithm \citep{HuffMcal2017, SheldonMcal2017} to measure galaxy ellipticities from noisy images. The raw images were processed by the DES Data Management (DESDM) team  \citep{Sevilla2011,Morganson2018,DES_DR1}.

\texttt{METACALIBRATION} provides an estimate of the shear field using a self-calibration framework that uses the data itself to correct for selection effects in the response of the estimate to shear. Inverse variance weights are assigned to galaxies. The DES Y3 shear catalogue has 100,204,026 objects, with a weighted $n_{\rm eff}=5.59$~galaxies~arcmin$^{-2}$. The \texttt{METACALIBRATION} self-correction accounts for most of the multiplicative bias, but there is a remaining multiplicative bias of $2$ to $3$ per cent \citep{y3-imagesims}. This multiplicative factor is left uncalibrated but is parameterized and its uncertainty accounted for in our inference framework.

The shear catalogue has also been tested for additive biases (e.g. due to point spread function residuals; see \citealt*{y3-shapecatalog}). The catalogue is characterized by a non-zero mean shear which is subtracted at the catalogue level before performing any analysis. The shear catalogue is divided into four tomographic bins, selected so as to have {roughly} equal number density.

The catalogue is used to create shear maps with a \healpix{} pixelization of \nside{512}. This relatively low resolution removes small scales that we cannot confidently model. This is tested and discussed further in section~\ref{sec:results}. The estimated value of the shear field in the map pixels is given by:
\begin{equation}
\label{eq:pixelvalue}
\gamma_{\rm obs}^{\nu} = \frac{\sum_{j}\epsilon_j^{\nu}w_{j}}{\bar{R}\sum_{j}w_j}, \,\, \nu=1,2,
\end{equation}
\noindent where $\nu$ refers to the two shear field components, $w_j$ is the per-galaxy inverse variance weight, $\bar{R}$ is the average \texttt{METACALIBRATION} response of the sample, and the summations are taken over the galaxies lying in a particular pixel.

\subsection{Simulation map raytracing}

For each simulation, lens planes $\delta_{\rm shell}(\hat{\boldsymbol{\rm n}}, \chi)$ are provided at $\sim 100$ redshifts from $z=49$ to $z=0.0$, equally spaced in proper time. The lens planes are provided as \healpix{} maps and are obtained from the raw number particle counts:
\begin{equation}
    \delta_{\rm shell}(\pindex{}, s) = \frac{n_{\textrm{part}}(\pindex{}, s)}{\langle n_{\textrm{part}}(\pindex{}, s) \rangle_{\phi}}-1 \ \ ,
\end{equation}
where $n_{\textrm{part}}(\pindex{}, s)$ is the number of particles in pixel $\pindex{}$ for shell $s$ and $\langle \rangle_{\phi}$ denotes an average over pixels. 

The lens planes are converted into convergence planes $\kappa_{\rm shell}(\pindex{}, \chi)$ under the Born approximation using the \hyperlink{https://github.com/NiallJeffrey/BornRaytrace}{BornRaytrace} code\footnote{\url{https://github.com/NiallJeffrey/BornRaytrace}}. The shear planes $\gamma_{\rm shell}(\hat{\boldsymbol{\rm n}}, \chi)$ are obtained from the convergence maps using Eq.~\ref{eq:kappagamma}, the inverse \cite{KaiserSquires} algorithm. We down-sample from the original resolution of \nside{2048} to \nside{512} (with pixel size $\approx$ 7.2 arcmin). 

These convergence $\kappa$ and shear $\gamma$ maps are the true shear and convergence fields in thin redshift shells. To generate mock lensing maps as they would be observed, we must a) integrate over a mock redshift distribution $n(z)$ (see Eq.~\ref{eq:kappadelta}), b) simulate the effect of intrinsic alignment of galaxies, and c) add the effect of galaxy shape noise and missing data (i.e. sky masks).

\subsection{Intrinsic alignments of galaxies}

We model the intrinsic alignment of galaxies using a density-weighted Non-Linear Alignment (NLA) model. 

Using the NLA model~\citep{hirata_seljak, Bridle2007}, we relate the convergence signal that would result from pure intrinsic alignments (with no lensing), $\kappa_{\textrm{IA}}$, linearly to the local density field:
\begin{equation}
\kappa_{\textrm{IA}} (\pindex{},z) = - A_{\textrm{IA}} C_1 \rho_{\textrm{crit}}  \frac{\Omega_M}{D(z)}  \Big( \frac{1+z}{1+z_0} \Big)^{\eta_{\textrm{IA}}}  \ \delta( \pindex{},z)
\end{equation}
in a pixel $\pindex{}$ and for some shell redshift $z$. We use the standard value of $z_0=0.62$ and set $C_1=5\times 10^{-14}M_{\odot}h^{-2}$Mpc$^2$~\citep[as per][]{Bridle2007}. 

The density-weighting in our forward model modulates the standard NLA model, because the source galaxies trace the underlying density field, and so are preferentially observed in higher density regions. This effect is the same as the clustering term
in the tidal-torque alignment (TATT) model for intrinsic alignments~\citep{blazek_2019}. The implementation of the source clustering effect is discussed in section~\ref{subsec:mockshearmaps}.

The amplitude of intrinsic alignments $A_{\textrm{IA}}$ and the redshift evolution parameter $\eta_{\textrm{IA}}$ are allowed to vary in our analysis as nuisance parameters. By sampling each of these parameters from a prior, they will be implicitly marginalized as part of our simulation-based inference procedure (see section~\ref{sec:sampling}). We choose the following (weakly informative) priors for these parameters:
\begin{itemize}
    \item $A_{\textrm{IA}} \sim \mathcal{U}(-3,3)$.
    \item $\eta_{\textrm{IA}} \sim \mathcal{U}(-5,5)$.
\end{itemize}

The $\kappa_{\textrm{IA}}$ maps are generated with the \hyperlink{https://github.com/NiallJeffrey/BornRaytrace}{BornRaytrace} code using the simulated overdensity maps $\delta$. From these we generate shear maps that contain only intrinsic alignment signal (i.e. no lensing).

\subsection{Realistic mock shear maps}\label{subsec:mockshearmaps} 

\subsubsection{Source clustering}
Due to the effect of clustering of source galaxies, known as \textit{source clustering} ~\citep[described and detected in][]{source_clustering}, it would be insufficient to assume a single galaxy redshift distribution $n(z)$ that is constant across the sky. Instead, our model $n(z,{\pindex{}})$ of the galaxy redshift distribution depends on sky position via an input \healpix{} pixel ${\pindex{}}$.

When constructing shear maps from observed data catalogues, each \healpix{} pixel is assigned an average shear, the average taken over all galaxies that are within that pixel and that are in the correct tomographic redshift bin. Since these source galaxies trace the underlying large-scale structure, higher-order correlations between the number of galaxies in pixels and the weak lensing signal encoded in the shear become important. 

In our forward model for generating mock data, we model a per-pixel redshift distribution via the sky-averaged redshift distribution $\bar{n}(z)$, then modulated by the density of galaxies:
\begin{equation}
    n(z) \propto \bar{n}(z) (1 + b_g \delta).
\end{equation}
\noindent The modulation factor assumes a linear galaxy biasing model with bias parameter $b_g$; here $\delta$ is the matter overdensity as before. This modulation is combined with an overall rescaling of the shape noise contribution to preserve the expected overall noise variance.  We follow the procedure of \cite{source_clustering}, which contains further details. 

\subsubsection{Shape noise \& mask}
This procedure also uses the randomly rotated shapes of the observed DES catalogue galaxies to implicitly generate the average intrinsic shapes in our mock observations, contributing shape noise to our mock shear maps.

The sky mask does not have to be treated separately; it is simply the set of pixels that contain no source galaxies. Because the same sky mask is present in both the mock simulated data and the observed DES data, it will be implicitly taken into account as part of our inference pipeline.

\subsubsection{Multiplicative shear bias}

To account for the residual errors in the shape measurement, we include a multiplicative shear bias in the forward model of the mock data. For a multiplicative shear bias $m$, associated with the particular tomographic bin, we rescale the shear by a factor of $1+m$.

Using the results from image simulations as presented in~\cite{y3-imagesims}, we use the following priors on $m$ for the various tomographic bins:
\begin{itemize}
    \item $m_1 \sim \mathcal{N} (-0.0063, 0.0091)$. 
    \item $m_2 \sim \mathcal{N} (-0.0198, 0.0078)$. 
    \item $m_3 \sim \mathcal{N} (-0.0241, 0.0076)$.  
    \item $m_4 \sim \mathcal{N} (-0.0369, 0.0076)$. 
\end{itemize}

\subsubsection{Photometric redshift uncertainty}

\begin{figure}
\includegraphics[width=0.9\columnwidth]{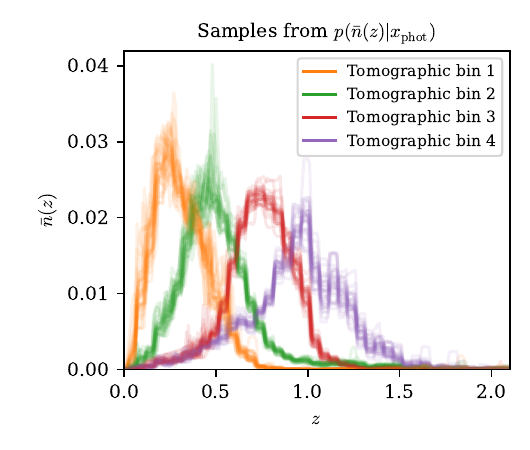}
\caption{An example set of samples from the \hyperrank{} probability distribution $p(\bar{n}(z) | x_{\rm phot})$ of the sky-averaged redshift distribution $\bar{n}(z)$ given the data used by \hyperrank{}. \label{fig:nz}}
\end{figure}

\begin{figure*}
  \includegraphics[width=1.05\textwidth]{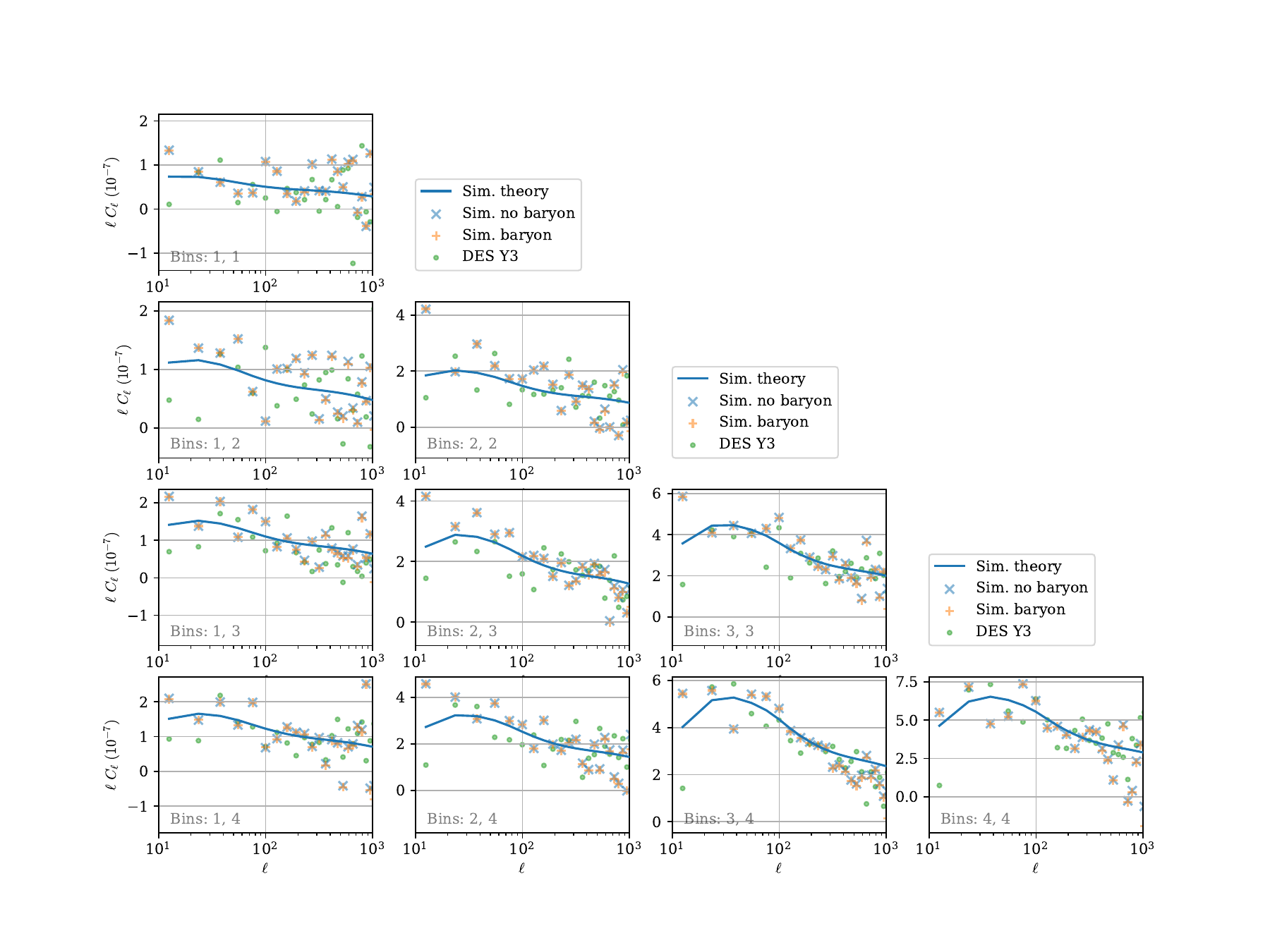}
 \vspace{-1.5cm}
 \caption{Power spectra $C_\ell$ (and cross-spectra between tomographic bins) of the baryonified simulation, the simulation without baryons, and the DES Y3 data. The theoretical power spectra, calculated with the same cosmological parameters as the two simulations, is shown for reference. \label{fig:cls}}
\end{figure*}

In the above discussion of source clustering, we used a fixed sky-averaged redshift distribution $\bar{n}(z)$. In fact, we sample realizations of possible distributions that are consistent with the data, based on the \hyperrank{} methodology~\citep[see][]{y3-hyperrank} using photometric redshift data $x_{\rm phot}$. 

The four tomographic bins are constructed to have roughly equal number density \citep*{y3-sompz} and the initial redshift distributions are provided by the SOMPZ method  \citep*{y3-sompz} in combination with clustering redshift constraints \citep*{y3-sourcewz} and correction due to the redshift-dependent effects of blending \citep{y3-imagesims}. Rather than using the best-guess $\bar{n}(z)$, \hyperrank{} generates realizations of possible $\bar{n}(z)$ samples in a way that marginalizes over redshift uncertainty.

Fig.~\ref{fig:nz} shows a random selection of $\bar{n}(z)$ samples. Each mock realization uses a different randomly sampled $\bar{n}(z)$; this contributes uncertainty in the photometric redshift distributions through our forward model, so that this uncertainty is taken into account in the inference pipeline.

\subsubsection{Mock shear map summary}

In summary (following \citealt{source_clustering}), the mock shear signal at \healpix{} pixel $\pindex{}$ in a thin simulated shell labelled with its redshift $z$ is generated according to

\begin{equation}
\begin{split}
\gamma({\pindex{}}) = \frac{\sum_z \bar{n}(z) [1 + b_g \delta({\pindex{}}, z)] (1 + m) [\gamma({\pindex{}}, z)+\gamma_{\textrm{IA}}({\pindex{}}, z)]}{\sum_z \bar{n}(z) [1 + b_g \delta({\pindex{}}, z)]} + \\
\left(\frac{\sum_z \bar{n}(z)}{\sum_z \bar{n}(z) \left[1 + b_g \delta({\pindex{}}, z)\right]}\right)^{1/2} F({\pindex{}}) \, \frac{\sum_g w_g e_g}{\sum_g w_g}.
\end{split}
\end{equation}
\noindent where $\bar{n}(z)$ is a \hyperrank{} sample that varies between each mock simulation.

The $F(\pindex{})$ factor provides the overall rescaling of the noise; we use $F(\pindex{}) = A (1-B \sigma_{e}^2(\pindex{}))^{1/2}$ where $A = [0.97,0.985,0.990,0.995]$ and $B = [0.1,0.05,0.035,0.035]$ for the four tomographic bins, and where $\sigma_{e}^2(\pindex{})$ is the shape noise pixel variance.

\section{Summary statistics \& compression} \label{sec:summarycompression} 

\subsection{Map making \& scale cuts}

The mock data is prepared using DES Y3 footprints in \healpix{} format, as described in \ref{subsec:mockshearmaps}. These shear maps are degraded to \nside{512}, corresponding to a scale cut of $6.9$ arcmin. Such a hard cut in pixel space corresponds to a smooth suppression of power in harmonic space (around 30 per cent by $\ell=1024$); for completeness, we also apply a hard cut at $\ell=1024$. 

The maps are converted from the shear fields to convergence fields using the Kaiser-Squires reconstruction, described by Eq.~\ref{eq:kappagamma}, where both E and B-mode convergence maps are retained.
  
\subsection{Power spectra, peaks \& neural compression}
\label{sec:powerpeakscompression}

\textbf{Power spectra}: The power spectrum $C(\ell)$ of a field on the celestial sphere is defined via

\begin{equation}
  \langle a_{\ell m}^{} a_{\ell' m'}^* \rangle =  C(\ell) \delta_{m m'} \delta_{\ell \ell'} \ .
\end{equation}

\noindent Here $a_{\ell m}$ are the spherical harmonic coefficients of the field, $\delta$ is the Kronecker delta, and the expectation $\langle \rangle$ is with respect to random realizations. An unbiased estimate of this power spectrum is
\begin{equation}
    \hat{C}(\ell) = \frac{1}{2 \ell +1} \sum_{m=-\ell}^{\ell} | a_{\ell m}^{}|^2 \ \ .
\end{equation}
It is the power spectrum of the shear field (not the convergence field) that we measure.  We decompose the shear field into $E$- and $B$-modes (curl-free and divergence-free components, respectively), yielding shear power spectra $C_\ell^{EE}$, $C_\ell^{EB}$, and $C_\ell^{BB}$. As with previous DES power spectra analysis, we use a pseudo-$C_\ell$ estimator that corrects for the effect of the sky mask. See \cite{Doux_2022} for details. This correction is not actually necessary to give unbiased results, as the correction (or lack thereof) would be applied equally to the simulated and observed data. Following the pseudo-$C_\ell$ correction, we obtain $C_\ell^{EE}$ and $C_\ell^{BB}$ to use as our observed data vectors. 

Fig.~\ref{fig:cls} shows the measured $C_\ell^{EE}-C_\ell^{BB}$ spectra for all tomographic bins, along with simulated spectra (discussed in section~\ref{sec:validation}).

\begin{figure*}
  \includegraphics[width=1.02\textwidth]{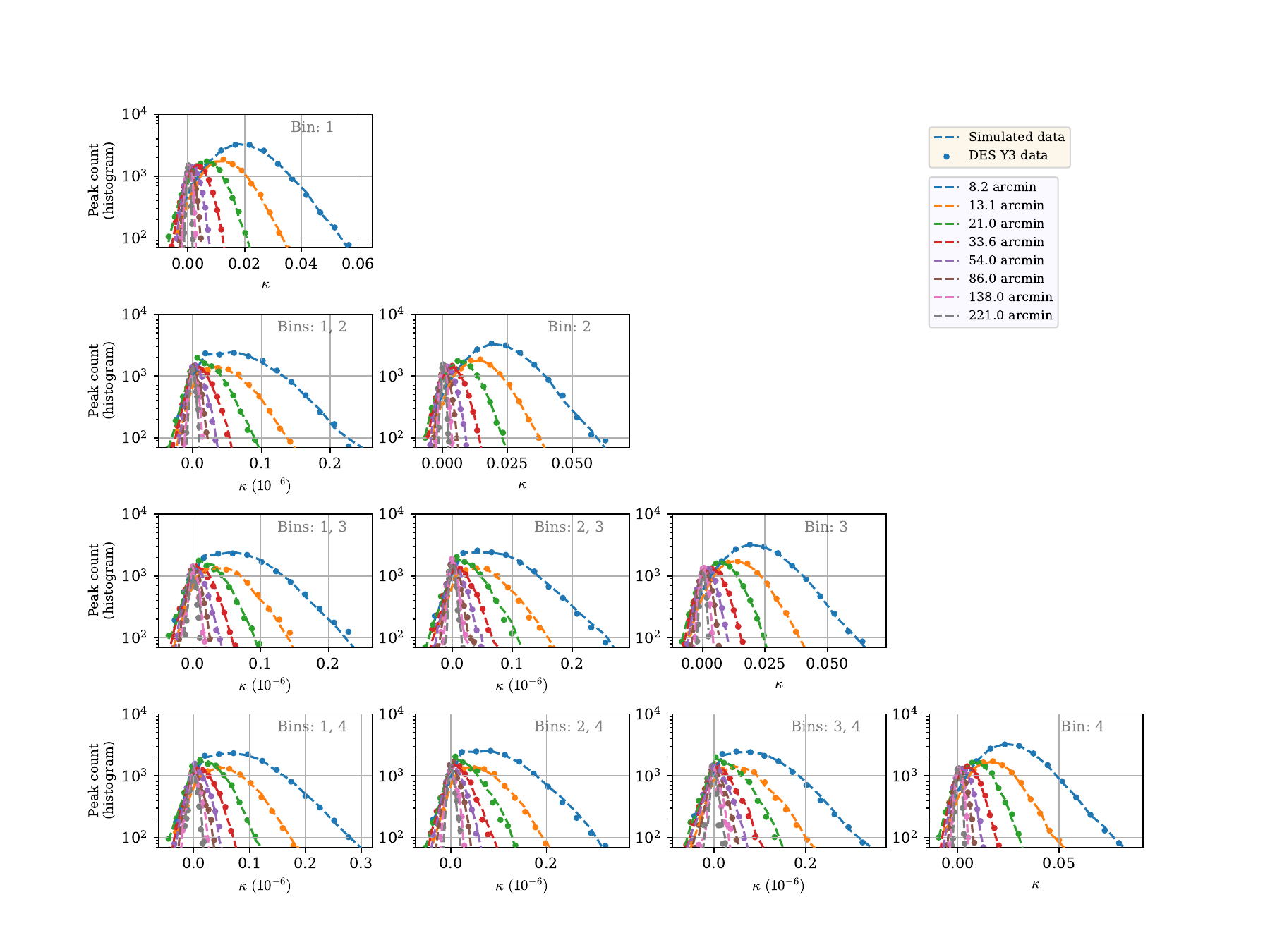}
 \vspace{-1.27cm}
 \caption{Peak count histograms from the DES Y3 data (solid circular markers) for each tomographic bin (rightmost plot on each row) and for the cross-maps. Each plot shows the observed peak counts for eight smoothing scales of the lensing map $\kappa$. For reference, we also show (dashed lines) histograms for a simulation that was chosen for its similarity with the actual observed data; it has $\Om=0.29$, $S_8=0.82$, $w=-0.83$, and randomly sampled nuisance parameter values. The cross maps have small $\kappa$ values because Eq.~\ref{eq:cross_maps} is not normalised.  \label{fig:peakdata}}
\end{figure*}

\noindent \textbf{Peaks:}  A \textit{peak} is a map pixel whose value exceeds that of its neighbouring pixels (typically there are eight such neighbours). For a given convergence map we create a histogram of the values of the convergence field at the peak pixels. Our histograms use 14 equally spaced bins and the range covered by these bins is chosen in advance so that each bin has at least ten peaks at a fiducial cosmology. We repeat this procedure on smoothed versions of our maps. This smoothing uses a top-hat filter; recall that in harmonic space the effect of such smoothing is to multiply the harmonic coefficients by
\begin{equation}
\label{eq:filter}
W_{\ell}(\theta_0) = \frac{P_{\ell-1}({\rm cos}(\theta_0))-P_{\ell+1}({\rm cos}(\theta_0))}{(2\ell+1)(1-{\rm cos}(\theta_0))},
\end{equation}
where $P_{\ell}$ is the Legendre polynomial of order $\ell$, $\theta_0$ the smoothing angle, and $\ell$ the multipole. We consider eight smoothing angles $\theta_0$ equally (logarithmically) spaced from $8.2$ to $221$ arcmin.
We count the peaks for the convergence maps from each of the four tomographic maps of the DES Y3 weak lensing sample. In addition, we account for cross-correlation between bins by following \cite{zuercherpeaks} and introducing `cross-maps' $\kappa^{ij}(\theta, \phi)$. These are new maps obtained by combining two original convergence maps for different tomographic bins $i$ and $j$ (with $i > j$):
\begin{equation}
    \label{eq:cross_maps}
    \kappa^{ij}(\theta, \phi) = \sum_{\ell = 0}^{\ell_{\mathrm{max}}}\sum_{m = -\ell}^{\ell} \hat{\kappa}^i_{ \ell m} \hat{\kappa}^j_{\ell m} Y^{}_{\ell m}(\theta, \phi),
\end{equation}
We compute the peak function in each of the resulting six new cross maps. Finally, before compressing the peak function, we adjust the peak counts of the noisy maps by subtracting the peak counts from a noise-only version of the maps.

\noindent \textbf{Compression:}  The power spectra compression uses an ensemble of 12 multi-layer perceptron (MLP) networks. As discussed in section~\ref{sec:compression}, we use an MSE loss function. All final fully-connected layer outputs use the sigmoid activation function; as a result, compressed statistics are confined to a sensible domain. This choice of activation function therefore requires a rescaling of our parameters (so that their prior ranges lie well within the bounds of the sigmoid activation).

Each MLP has an input size of $560 \, (= 10 \times 28 \times 2)$, corresponding to the ten cross correlations of the four tomographic bins, in $28$ multipole ($\ell$) bins, over the two components (EE and BB) of shear maps. The MLP network has ten hidden layers, each with 256 nodes, with an embedded layer normalisation and a ReLU activation function at each layer output. The last layer reduces the output size to a single node corresponding to the selected parameter being compressed. Similarly to the CNN ensemble, there is a final sigmoid activation function on the output of the final layer. This MLP network is trained 12 times, each using the same data but different random network parameter initializations, and the resulting 12 predictions are averaged to yield an ensemble prediction. (We trained 12 times because this was clearly superior to training just once, and because it was convenient for the computer hardware being used; we do not claim optimality.) 

The training input data is augmented using additive random Gaussian noise as a regularization measure. The noise added to each input bin $\ell_i$ is sampled from  $\mathcal{N}(\mu_i,\sigma_i \times 10^{-3})$, where $\mu_i$ and $\sigma_i$ are the mean and standard deviation of the values in each multipole bin $\ell_i$ observed across the training dataset. Each network is optimized using the stochastic gradient decent Adam's optimizer with a mean squared error (MSE) as the loss metric. The learning rate was initially set at $1\times10^{-4}$ and decays exponentially with a decay rate of $0.1$ per training step. The training is performed for up to 200 epochs with an early stopping criterion; this is described in more detail in \ref{subsec:CNNcompression}.

Compression of the peak counts is done similarly; the only difference is that the MLP has an input size of $1120 \, (= 10 \times 14 \times 8)$.

\begin{figure*}
  \includegraphics[width=0.8\textwidth]{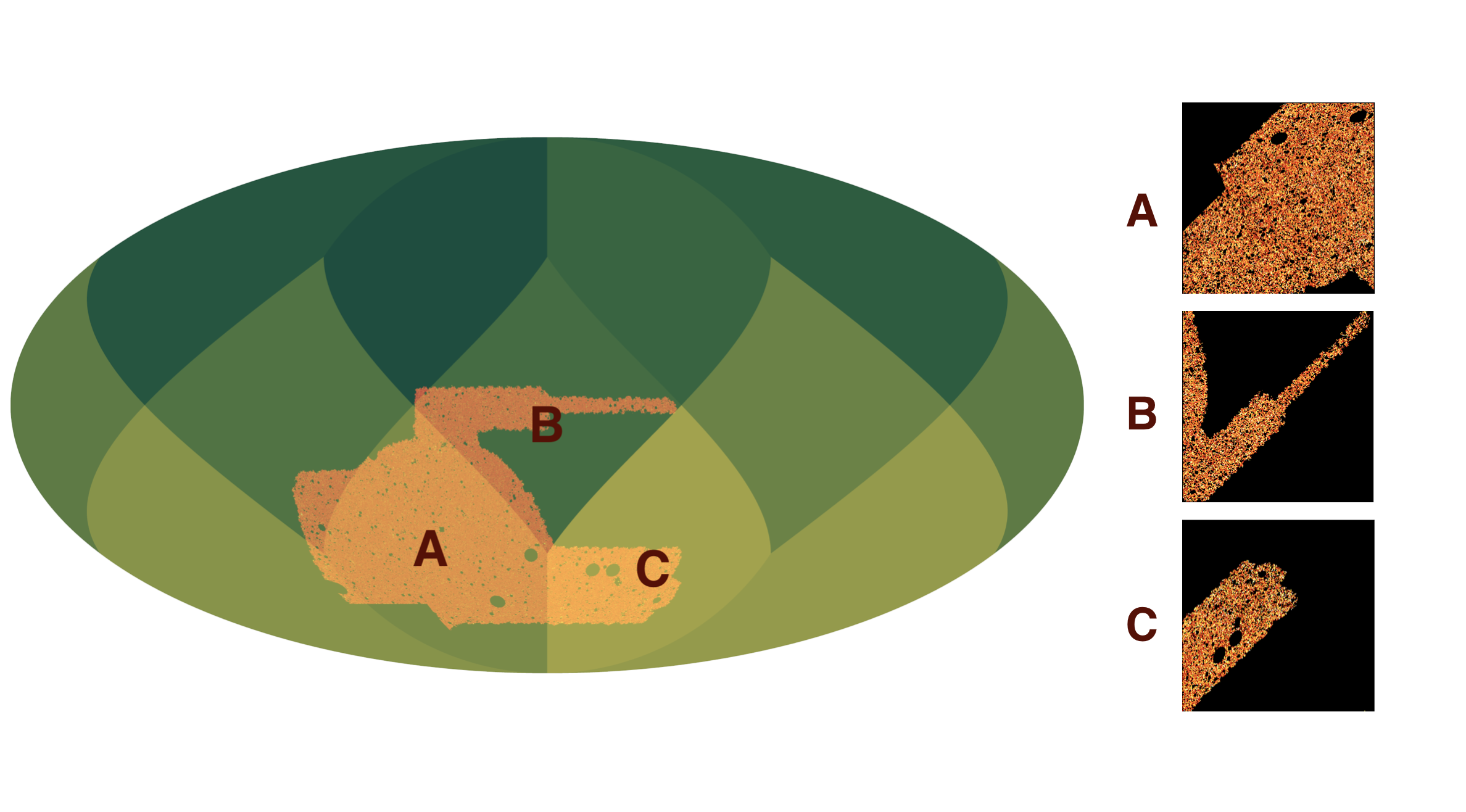}
 \vspace{-0.8cm}
 \caption{A demonstration of the CNN patching scheme for map-level compression, showing an example mock convergence ($\kappa$) \healpix{} map (in orange) with \nside{512}. The pixels of the convergence map are split into patches $A, B, C$ based on the respective \nside{1} pixel (in green). This produces the square patches seen to the right of the figure to be that are used to train the CNN ensemble. This patching scheme is lossless: there is a one-to-one match between the \healpix{} map pixels and the patch pixels.  \label{fig:patchescutout}}
\end{figure*}
\subsection{Map-level (CNN) compression}\label{subsec:CNNcompression}

This approach aims to infer cosmology directly from the map data. Here we implement a Convolutional Neural Network (CNN) as a higher order statistic, using deep learning to compress relevant features directly from pixels; CNNs can be optimized to compress these features to a lower dimension in an informative way. We make no claim that our map-level compression is optimal (in the sense that the resulting parameter constraints are the best possible), but it is practical and does lead to significantly improved results.

Planar CNNs take flat two-dimensional images as input whereas our data (via the \healpix{} pixelization) are embedded on the sphere. There exist neural networks adapted to the geometry of the sphere \citep[e.g.][]{deepsphere_iclr, DISCO}. For practicality, however, we have decided instead to perform separate analyses of several nearly-flat rectangular patches on the sky. By using patches we lose large-angle correlation information, but this can be mitigated by combining (via concatenation of compressed data vectors) the CNN output with the compressed power spectrum output (as described in section~\ref{sec:summarycompression}) -- on large-angle scales we expect the signal to be near-Gaussian (and hence for these scales the power spectra are already maximally informative). We will refer to this combination as \CLCNN{}.

For our CNN approach we use patches from the sphere, flattened to two-dimensional images of $512 \times 512$ pixels. The `nested' format for \healpix{} pixel ordering offers a natural method for extracting new patches; we form a patch by taking all the \nside{512} resolution pixels that lie within a single superpixel defined by the minimum \healpix{} resolution \nside{1}, as shown in Fig.~\ref{fig:patchescutout}. This projection distorts the spherical geometry, which for traditional parameter estimation approaches would bias the inference. However, in our simulation-based method this transformation will also be applied consistently to the DES Y3 data, and therefore projection distortion will not bias any parameter estimation. Nevertheless, projection distortion makes the compression potentially suboptimal. Our chosen patch size is a compromise between loss of large-scale information (from having small patches) and projection distortion (from having large patches). The chosen scheme leads to the DES footprint being split into three patches (labelled $A,B,C$; a small subsection of the footprint is discarded). 

To complement this patching scheme we construct an ensemble of weak learning CNNs. Each patch has four dedicated networks, trained on the same data but with different network parameter initialization. Compressing each patch individually has the advantage of allowing each network to become familiar with the footprint of each patch. The resulting compression will be the weighted average of all 12 CNN networks (four per patch for three patches) for a single chosen parameter (with weights given by the sky fraction of each patch -- see  Fig.~\ref{fig:patchescutout}). This ensemble approach of averaging many simple CNNs has been shown to be a robust way to train networks that generalise well with smaller datasets to avoid over-fitting;

such an approach is advantageous considering the computational expense of constructing mocks.

The network is fed eight channels (E- and B-modes of the convergence maps for each of four tomographic bins); each channel supplies a patch of $512 \times 512$ pixels. The input maps are augmented during training by the same procedure described previously in \ref{sec:powerpeakscompression}, where the means and standard deviations are calculated for each individual pixel. The same setup for loss metric and optimisation function are applied identically to the CNN networks as in \ref{sec:powerpeakscompression}.

The ensemble CNN was trained on $9264$ DES Y3 mock data sets (three independent noise realizations of our 3088 original simulated convergence maps; see section~\ref{subsec:mockshearmaps}). Each individual CNN network in the ensemble was trained for up to 200 epochs, with early stopping to prevent over-fitting. The stopping criterion is based on the MSE loss of a set of 3088 data with a different noise realisation, which acts as a validation data set. We also use this validation data set when performing the neural density estimation task and this dual use of the validation data set has the potential to introduce bias. We have ruled out this possibility by checking that our inferred posteriors do not shift when using an additional different data set (to which the CNN is entirely blind during training) in place of the validation data when performing the neural density estimation task. This procedure attempts to mitigate any over-fitting in two different ways.

The ensemble CNN trains in just under one hour per parameter on 12 Nvidia A100 GPUs using the NERSC Perlmutter cluster.

As a final step in the algorithm, the CNN output and the compressed power spectrum data vector are concatenated.

\begin{table}
\caption {Prior and hierarchical probability distributions.}
\centering
\begin{tabular}{|c c|}
\hline
\textbf{Parameter} & \textbf{Prior probability distribution} \\
\hline 
$\Om$ & $\mathcal{U}(0.15,0.52)$ \\
$S_8$ & $\mathcal{U}(0.5,1.0)$ \\ 
$w$ & $\mathcal{U}(-1,\frac{1}{3})$  \\
\hline
$n_s$ & $\mathcal{N}(0.9649, 0.0063)$   \\
$h$ & $\mathcal{N}(0.7022, 0.0245)$ \\
$\Omega_{\rm b}h^2 $&  $ N(0.02237, 0.00015)$\\
$\log(m_{\nu}) $& $ \mathcal{U}[\log(0.06), \log(0.14)]$ \\
\hline
$A_{\textrm{IA}}$ & $\mathcal{U}[-3, 3]$  \\
$\eta_{\textrm{IA}}$ & $\mathcal{U}[-5, 5]$ \\
$m_{1}$ & $\mathcal{N}(-0.0063,0.0091)$ \\
$m_{2}$ & $\mathcal{N}( -0.0198,0.0078)$ \\
$m_{3}$ & $\mathcal{N}( -0.0241,0.0076)$ \\
$m_{4}$ & $\mathcal{N}(-0.0369, 0.0076)$ \\
$\bar{n}_i(z)$ & $p_{\hyperrank{}}(\bar{n}_i(z) | x_{\rm phot})$  \\ 
\hline
\end{tabular}
\label{tab:priors}
\end{table}

\section{Results}\label{sec:results} 

\subsection{Prior probabilities}

Table~\ref{tab:priors} summarizes the priors used in our cosmological inference.

The first (top) group includes the three target parameters. In our mock data these parameters are \textit{not} distributed according to our chosen prior. The parameters $S_8$ and $\Om$ were sampled with active learning and have a particularly strange distribution in Fig.~\ref{fig:gower_street_params}. The prior, therefore, must always be set explicitly when combining with the learned likelihood; furthermore, we must always include these parameters in our learned likelihood (unlike the parameters with implicit priors described below).

The middle group of parameters in Table~\ref{tab:priors} have implicit priors, i.e. the Gower Street simulations have these parameter distributions (see section~\ref{sec:gowersims} for caveats). Even if these parameters are not explicitly included in the learned likelihood, they will be implicitly marginalized during inference and the uncertainty in these parameters will be propagated to the final constraints on other parameters (see section~\ref{sec:sampling}).

The final (bottom) group in Table~\ref{tab:priors} are the nuisance parameters, which are varied according to these prior distributions during mock shear map generation. Again, the uncertainty in these parameters is propagated through forward modelling in this simulation-based inference framework.

\begin{figure}
\includegraphics[width=0.9\columnwidth]{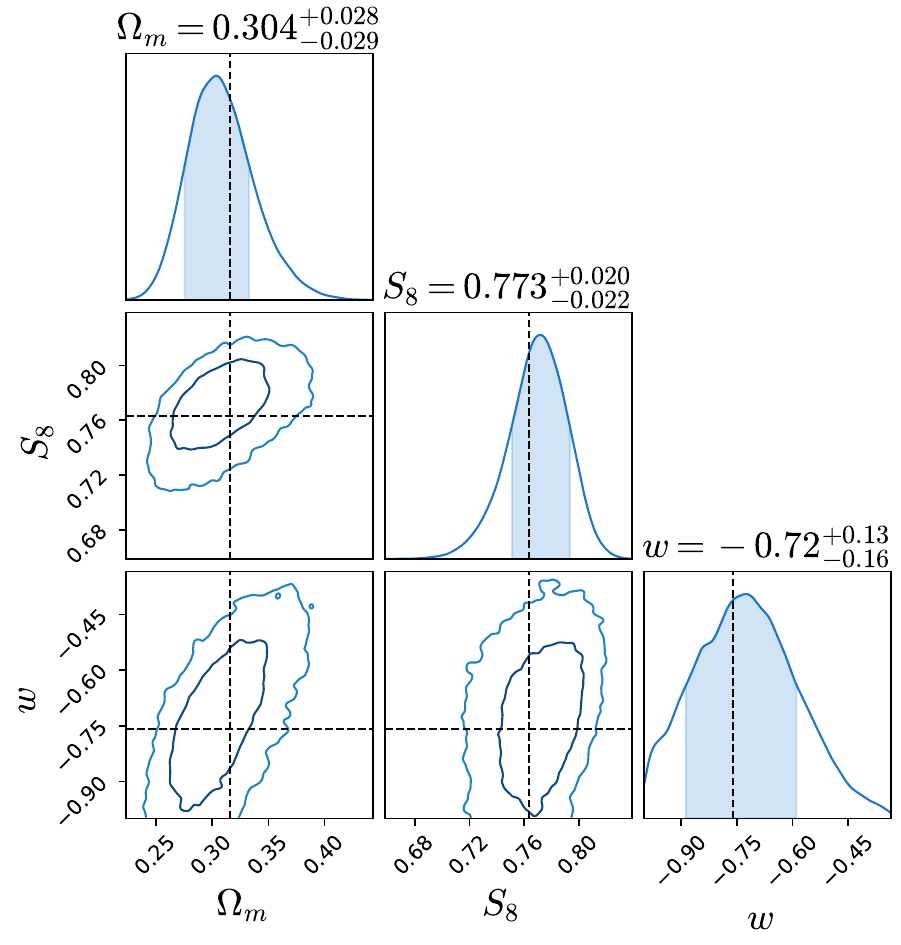}
\caption{Inference with the mean mock data (combining mock power spectra and map-level CNN compressed data). The values for $S_8$ and $\Om$ denoted by dashed lines are the average of the true parameter values from the same mock data. This is analogous to using noise-free data for inference when using a Gaussian likelihood. \label{fig:cnn_noisefree_mock}}
\end{figure}

\subsection{Simulation validation with Gower Street sims}

\subsubsection{Inference from mean mock data}

We test that we can recover the correct `input' parameter values from averaged data. This test is an analogue of the standard `noise-free' inference, which is typically performed to demonstrate recovery of the input parameters. Instead, we take the average data vector from a set of mock simulations, perform inference, and validate that we recover the average parameter values from the same set of mock simulations.

The data vector used for this inference is the per-element mean of all compressed mock data vectors: $\bar{t}_j = \frac{1}{N} \sum_{i=0}^{N-1} ({t_j})_i$ where the $i$ index denotes individual mock data vectors (over the full parameter space for the Gower Street sims) and $j$ indexes the elements of the data vector. This mean data vector uses all mock compressed data vectors that were used for the density estimation. 

Fig.~\ref{fig:cnn_noisefree_mock} shows the result of this test for the combination of mock power spectra $C_\ell$ and the map-level (CNN) compression. The mean parameter values are clearly recovered. We can confirm this test is also passed for the other combinations of data.

\subsubsection{Coverage test results}

As introduced and described in section~\ref{sec:coverage_test_theory}, coverage tests repeat the parameter inference procedure to test that the estimated posterior describes the correct probability for the parameters. We repeat the inference procedure, each time excluding one mock data vector from the neural likelihood estimation step. The likelihood is then evaluated using that held-out data vector, with the posterior then being evaluated and compared to the true parameter value.

We use the \tarp{} package to estimate the coverage probabilities in the three-dimensional parameter space $\{\Om, S_8, w\}$ (rather than on the marginal posteriors individually). This code implements the `Tests of Accuracy with Random Points' (TARP) algorithm, which estimates coverage probabilities of generative posterior estimators. We repeat the neural likelihood estimation technique 50 times, we draw $9\times 10^3$ samples from the learned posterior conditioned on the held-out data vector, and we perform coverage testing using these Markov chain Monte Carlo (MCMC) samples as an input to \tarp{}.

\begin{figure}
\includegraphics[width=0.85\columnwidth]{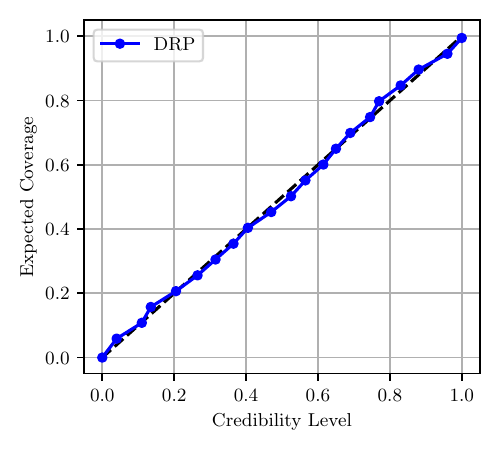}
\caption{ Coverage test result (using \tarp{}) to validate the inference pipeline. Using repeated mock data parameter inference, the fraction of true values in the appropriate credible intervals matches the expected fraction. The figure shows the result for the map-level CNN patch compression; similar coverage tests were successful for the other observables. DRP is the `Distance to Random Point' ~\citep[see][]{lemos2023sampling}. \label{fig:tarp} }
\end{figure}

Fig.~\ref{fig:tarp} shows the result of this procedure applied to the map-level CNN patch data (plots for the other observables show similar results). The expected coverage does indeed match the credibility level. This validates our neural likelihood estimation, showing that the posterior distribution (i.e. parameter uncertainties) truly represent the probabilities that the Universe has some true parameter value.

\subsection{Robustness to mismodelling \& residual systematic errors}

\subsubsection{Systematic error injection}

We describe tests each of which confirms that the variation of some source of systematic error does not affect our results. In the language of machine learning statistics, these are robustness tests for (a specific type of) \textit{distributional shift} — a mismatch between the training data and the deployment data.

The tests use the \texttt{CosmoGridV1} simulations suite~\citep{cosmogrid}. We chose a set of one hundred simulations at the fiducial cosmology $\sigma_8 = 0.84$, $\Omega_{\rm m}=0.26$, $w=-1$, $H_0=67.36$, $\Omega_{\rm b}=0.0493$, $n_{\rm s}=0.9649$. The \texttt{CosmoGridV1} simulations, like the Gower Street simulations, were created using the \pkdgrav{} code \citep{potter2017pkdgrav3}; they were created independently, however, and hence can serve as a further test that the correct input cosmology is recovered.

For each source of systematic error we generate two sets of mock data with different levels of systematic error included. We then apply our inference pipeline to each of these mock data sets and compare the resulting posterior probability distributions of the cosmological parameters.

To test for any overall shifts in the resulting posterior distributions, we use the average compressed data as the input to the neural likelihood estimation. For example, for the fiducial result, we measure each summary statistic for each of our selected mock \texttt{CosmoGridV1} data sets, then compress each separately, then average the resulting compressed statistic. This averaging mitigates the intrinsic variability that is expected between different data realizations.

Two sources of systematic error are tested: baryon feedback and source clustering bias variation.
\begin{itemize}

\item {
    \noindent \textit{Baryonic feedback} – Feedback effects can lead to suppression of structure on small cosmological scales. The Gower Street sims are N-body (dark matter only) simulations and do not include any baryonic astrophysics. In line with the standard DES weak lensing analyses, we cut scales that we think are likely to be affected by baryons and then test for the effect of possible baryonic contamination (e.g. \citealt{y3-cosmicshear1}; \citealt*{y3-cosmicshear2}; \citealt{zuercherpeaks}).

    Hydrodynamical simulations are unfortunately too computationally expensive to generate a sufficient quantity of realistic mock data that include baryons. We therefore use the \texttt{CosmoGridV1} maps, as these include a baryon correction model; this model \citep{cosmogrid} changes the density fields (in a post-processing step) to emulate baryon feedback.

    The effect of baryons on the simulated power spectra can be seen in Fig.~\ref{fig:cls}, which shows the measured power spectra from data in the fiducial set (`Sim. no baryon') and from data with baryon feedback included (`Sim. baryon'). The suppression can be seen at small scales (high $\ell$).
}
\item{
    \noindent \textit{Source clustering bias variation}  – Although we expect the effect to be small, it is possible that a different value for galaxy bias of the source galaxies could change our results. This is due to source galaxy clustering, which is known to change the predicted observations (section~\ref{subsec:mockshearmaps}). We use a fixed value of galaxy bias $b=1$ in our forward model, and hence we need to test that our results are not sensitive to a different true value in the observed data.

    We generated two sets of simulated mock data, in addition to our fiducial mock data: one with a high galaxy bias ($b=1.5$) and one with a low galaxy bias ($b=0.5$). 
}

\end{itemize}

\begin{figure}
\includegraphics[width=1.02\columnwidth]{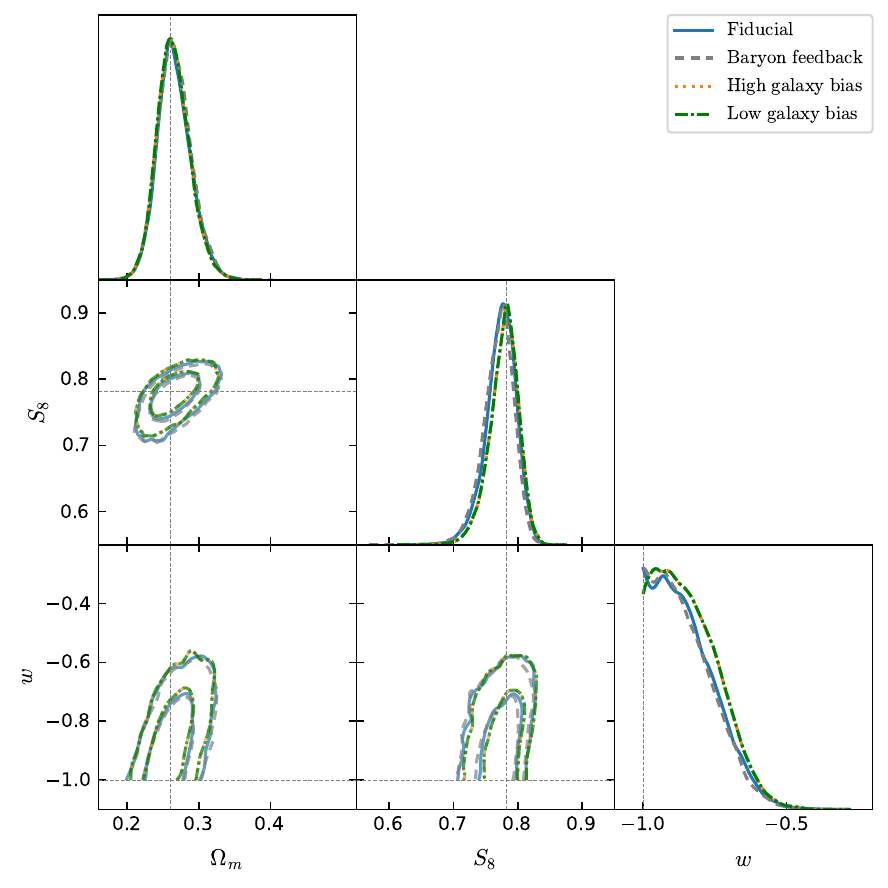}
\caption{The marginal posterior distributions with independent \texttt{CosmoGridV1} simulated data with two sources of systematic error in the mock data: baryon feedback and varying source galaxy bias. Both of these are found to induce changes in the posterior below 0.3 $\sigma$ in the marginal $S_8$-$\Om$ plane (the standard Dark Energy Survey test). This test also shows that our pipeline recovers the true parameter values (straight dashed lines) with independent mock data. The shifts in the mean of marginal posterior distribution of $w$ are all below 0.3$\sigma$. \label{fig:syst_plot}}
\end{figure}

Test results are shown in Fig.~\ref{fig:syst_plot}, which plots the posterior distribution for \CLCNN{} (power spectrum combined with map-level compression). Each of the two effects tested has a relatively small impact on the posterior. For the change in the $S_8$-$\Om$ marginal posterior we use the standard DES criterion: we measure the shift in marginal posterior distribution relative to the standard deviation, finding that each of these systematic effects induces a shift below 0.3$\sigma$ (the maximum level set by DES analyses). This test is also passed for $C_\ell$ (power spectra alone) and for \CLPEAKS{} (power spectrum combined with peak counts).

Note that the true value of the input data $w=-1$ is on the boundary of the prior. We test the shift of the mean of the marginal posterior for $w$, finding that both systematic effects induce a shift of less than 0.3$\sigma$.

\subsubsection{Further systematic errors in lensing maps}

A potential source of contamination in the observed data is the misestimation of the point spread function (PSF). Failures in PSF modelling can cause errors in the measured shapes of galaxies, characterized by an additional ellipticity component $\delta \epsilon^{\textrm{sys}}_{\textrm{PSF}}$.

\cite{Jarvis2016} and \cite*{y3-shapecatalog} provide a model to describe $\delta \epsilon^{\textrm{sys}}_{\textrm{PSF}}$ that can be calibrated using \textit{reserved stars}, i.e. those stars not used to train the original PSF model. We therefore could, following the procedure described in~\cite{gatti2023dark}, generate a map of $\delta \epsilon^{\textrm{sys}}_{\textrm{PSF}}$ per tomographic bin to be added to the fiducial shear maps; this could serve as a high, but in principle possible, contamination due to PSF errors. However, if we inject this PSF contamination at map level, we are overwhelmed by shot noise from our finite sample of reserved stars and this particularly affects small scales. We therefore do not use this approach (while we await further work that could accurately forward model PSF errors into our mock lensing maps).

We instead rely on alternative tests. In \citealt*{y3-shapecatalog}, the DES Y3 shear catalogue tests showed no evidence of additive biases due to PSF mismodelling. Furthermore, tests of reconstructed mass maps in~\cite*{y3-massmapping} showed no evidence of PSF residual errors.

\begin{figure}
\hspace{1.5cm}
\includegraphics[width=0.7\columnwidth]{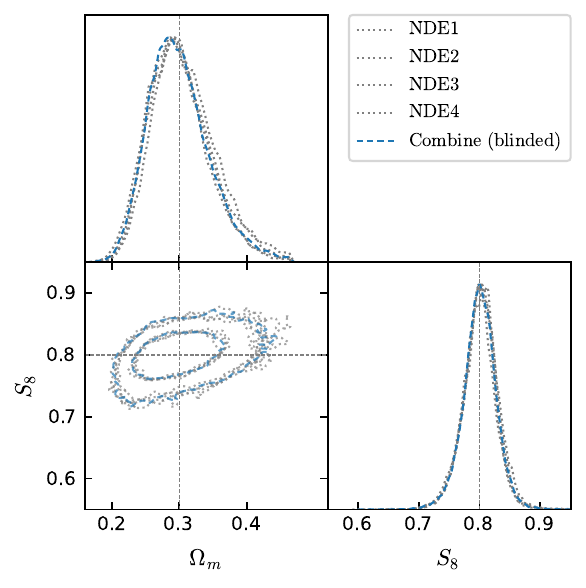}
\caption{Neural density estimator ensemble convergence test using power spectrum $C_\ell$ data. Although using the real observed DES Y3 data, this was a blind test, which was achieved by shifting the posterior mean to a fiducial value ($\Om=0.3$ and $S_8=0.8$).  \label{fig:cnn_test_ensemble}}
\end{figure}

\subsection{Blinded data likelihood ensemble validation}
\label{ssec:ensemblevalidation}
To test the convergence of the neural density estimation (i.e. the likelihood learned from simulated data), we compare the different density estimates that comprise the ensemble. As described in section~\ref{sec:ndeconvergence}, an insufficient number of simulated data realizations typically leads to significant differences between the neural likelihood estimates. With more simulations, the predictions from the ensemble converge.

Unlike the previous test on simulations, this test can be applied to the estimated likelihood evaluated for the actual observed data. This test was therefore done blind, as described in section~\ref{ssec:blindingstrategy}, and we confirmed that the test was passed before the full unblinded results were seen (section~\ref{subsec:unblindedresults}).

Fig.~\ref{fig:cnn_test_ensemble} shows the posterior distributions from each likelihood in the ensemble for the observed DES data. In this example the data are the power spectra $C_\ell$. This test was performed (and passed) before unblinding any of our results on data (including peaks and CNN map-level inference).  
 
\subsection{Blinding Strategy}
\label{ssec:blindingstrategy}
We used a blinding strategy (described below) to reduce the impact of confirmation bias. Blinding has been used by many DES analyses \citep{Muir_2020}, but note that the approach of this paper made necessary some deviations from the standard DES blinding strategy.
\begin{itemize}
\item{
Some simulations used input cosmological parameters obtained from `active learning' (see section~\ref{ssec:cosmoparams}), and this required estimating the posterior distribution of the cosmological parameters using the simulations available at that point. These estimations were held within the computer code and were not revealed to the experimenters. 
}
\item{
All training of the neural networks for compression and for density estimation was finalized without evaluation on any real observed data. The entire pipeline was run using a simulation (as if it were real data); results were checked for reasonableness and the neural network parameters were then frozen.
}
\item{
The uncompressed statistics from real data (the measured power spectra and peak counts) were checked for reasonableness.
}
\item{
The compressed statistics from real data were confirmed to be well within the convex hull of the scatterplot of compressed statistics obtained from the simulations.
}
\item{
The posterior distribution of cosmological parameters was inferred from observed data; this posterior was then shifted (by an amount that was kept within the computer code and was not available to the experimenters) to have a fiducial mean value. This shifted posterior was used in the likelihood ensemble validation of section~\ref{ssec:ensemblevalidation}. It was also used to confirm that the posterior distribution had a figure of merit similar to that derived in a similar way from simulations (note that the figure of merit is sensitive to the width of the posterior but not to its mean).
}
\item{
Finally the shift to the posterior mean was removed, revealing the unblinded posterior. 
}
\end{itemize}

\begin{figure}
\hspace{-0.2cm}
\includegraphics[width=1.02\columnwidth]{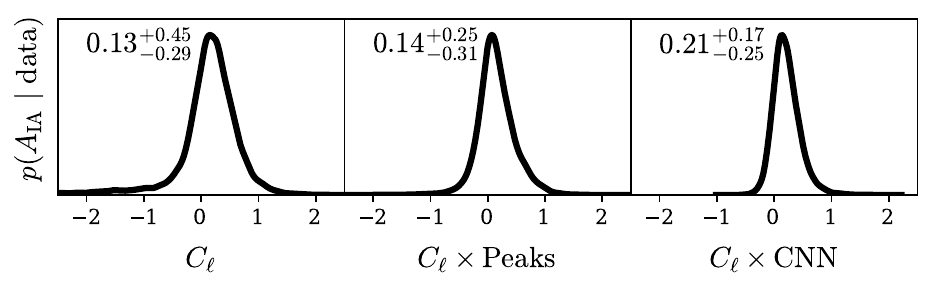}
\caption{Marginal posterior distribution of the amplitude of intrinsic alignment $A_{\rm IA}$ for the three DES Y3 combinations: power spectra $C_\ell$, peaks with power spectra \CLPEAKS{}, and map-level inference \CLCNN{}. These constraints use additional data and a different $w$ prior compared to our main cosmological results (see section~\ref{sec:ia_result} for discussion), to show that the inferred intrinsic alignments are reasonable from this analysis. \label{fig:ia_result}}
\end{figure} 

\subsection{Intrinsic alignments} \label{sec:ia_result}

\subsubsection{Discussion}

Our $S_8$ compression is not optimal. We find that including an additional compression of the map to informative summaries of $A_{\rm IA}$ (the intrinsic alignment amplitude) improves our posterior constraints on $S_8$. This shows that the original $S_8$ compression was missing information; this was to some extent expected (see section~\ref{subsec:CNNcompression} for a discussion).

This does not mean the intrinsic alignment nuisance parameters are incorrectly marginalised when using only the sub-optimal $S_8$ summary, $t_{S_8}$. That is, the following desired property still holds:
\begin{equation}
p(t_{S_8} | S_8) = \int p(t_{S_8} | S_8, A_{\rm IA}) \ p(A_{\rm IA}) \ {\rm d} A_{\rm IA}  \  \ ,
\end{equation}
\noindent where $p(t_{S_8} | S_8)$ is a learned likelihood for $S_8$. What \textit{is} true is that the posterior $p(S_8 | t_{S_8},  t_{A_{\rm IA}})$ is tighter than $p(S_8 | t_{S_8} )$; this is because $t_{S_8}$ is a sub-optimal summary statistic.

Despite the possibility of improved cosmological constraints, we nevertheless do not include in our main analysis the compressed $A_{\rm IA}$ summary statistic, $t_{A_{\rm IA}}$. The primary reason for this choice is that including this statistic results in a posterior distribution so tight that the NDE ensemble test fails; the density of simulations in that region of parameter space becomes too low.

Even if this test had not failed, there would be further reasons to not include this additional information. The intrinsic alignment NLA model has been tested with direct two-point correlation measurements only down to scales of $\sim 5 {h^{-1}\rm Mpc}$~\cite[e.g.][]{Johnston_2019, SinghIA}; at smaller scales linear galaxy bias modelling is insufficient. At the peak of our redshift distribution of source galaxies, around $z \sim 0.6$, our angular scale cuts correspond to a physical scale of $\sim 3 {h^{-1}\rm Mpc}$. We may have some confidence that the NLA model continues to hold at such small scales (unless there are unexpected higher-order contributions); nevertheless, if the constraints on $S_8$ are strongly affected by $t_{A_{\rm IA}}$ then it is prudent to exclude this additional information.

\subsubsection{Results}

Although we do not include the $t_{A_{\rm IA}}$ compressed statistics in our analysis for cosmological constraints (section~\ref{subsec:unblindedresults}), here we present the marginal posteriors for $A_{\rm IA}$ using the $t_{A_{\rm IA}}$ statistics as a `sanity check', i.e. to confirm that the inferred $A_{\rm IA}$ values are reasonable.

Fig.~\ref{fig:ia_result} shows the marginal posteriors for the intrinsic alignment amplitude $A_{\rm IA}$ for each of our standard data combinations: power spectra, peaks with power spectra, and map-level inference. To reduce the NDE dimension, we implicitly marginalize $w$, so that the $w$ prior is given by $\mathcal{N}(-1,1/3)$ for values of $-1<w<-1/3$ (see section~\ref{sec:sampling} for discussion). This change does not particularly impact intrinsic alignment inference, and, furthermore, the aim of this inference is just to confirm that the $A_{\rm IA}$ values are reasonable.

The results for $A_{\rm IA}$ are all consistent, with a slight preference for low positive values, but still consistent with $A_{\rm IA}=0$. This result is also consistent with the results from existing DES two-point analyses \citep[e.g.][]{y3-cosmicshear1,y3-cosmicshear2, Doux_2022}.

\subsection{DES Y3: Cosmological constraints}
\label{subsec:unblindedresults}
We present results using the three data combinations previously described: power spectra ($C_\ell$), peaks and power spectra (\CLPEAKS{}), and map-level inference (\CLCNN{}).

As described in section~\ref{sec:compression}, these data (summary statistics) $x$ are compressed to lower-dimensional summary statistics $t=\mathcal{F}(x)$. In each case the compression function $\mathcal{F}$ is a neural network (or ensemble of networks), optimized for the given summary statistic.

We target the parameters $\Om$, $S_8 (\equiv \sigma_8 (\Om /0.3)^{1/2}$), and $w$, and thus the compressed data for a given summary statistic has three elements. When we combine the data (e.g. \CLPEAKS{}) we concatenate the compressed data vectors (e.g. $t = {\mathrm{concat}}[t_{C_\ell}, t_{\rm Peaks}]$), giving a compressed data vector with six elements. All other parameters, including cosmological and  nuisance parameters, are implicitly (and correctly) marginalized; see section~\ref{sec:sampling} for details.

\begin{figure}
\includegraphics[width=1.02\columnwidth]{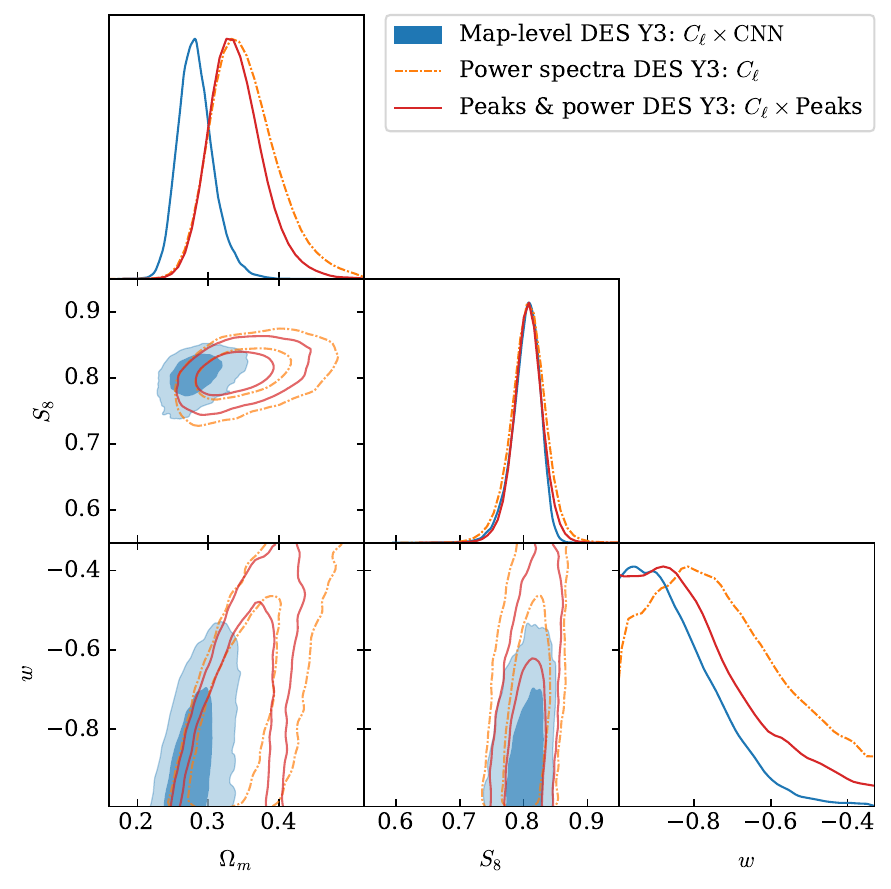}
\caption{Posterior probability distribution for $\{\Om, S_8, w\}$ obtained from simulation-based inference using three DES Y3 data combinations: power spectra $C_\ell$, peaks with power spectra \CLPEAKS{}, and map-level inference \CLCNN{}. \label{fig:results_data}}
\end{figure}

Fig.~\ref{fig:results_data} shows the marginal two-dimensional posterior distribution for the three data combinations. Credible intervals derived from the one-dimensional marginals were calculated using the \textsc{GetDist} package~\citep{lewis2019getdist} and are listed in Table~\ref{tab:results}. This figure and table show the main result of this paper.

All of these results use a full simulation-based (likelihood-free) inference pipeline to infer cosmological parameters.

We can also compare our most constraining result, that from \CLCNN{} (map-level inference combined with power spectrum), to existing data and likelihoods. We compare to the Planck cosmic microwave background (CMB) data; here we use the Planck likelihood code with model priors amended to match our analysis choices (except for $\Omega_{\rm b}$, as our prior here was motivated by Planck). We also compare to the DES Y3 likelihood for real-space weak lensing two-point correlation functions; here also we have matched the analysis choices and priors (where appropriate).

\begin{figure}
\includegraphics[width=1.02\columnwidth]{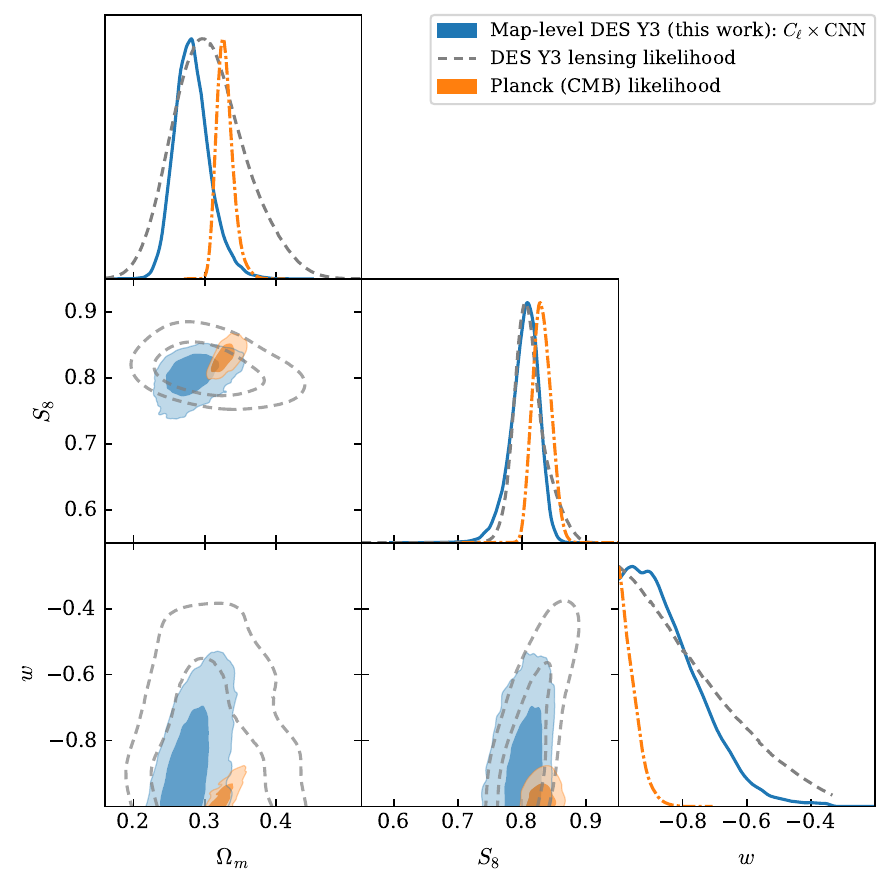}
\caption{Comparison of the \CLCNN{} result (map-level compression) with results both from the Planck CMB likelihood and from the standard DES weak gravitational lensing (two-point correlation function) likelihood (both of which have been subject to reanalysis to match prior choices).
\label{fig:results_data_comparison}}
\end{figure}

Fig.~\ref{fig:results_data_comparison} compares our analysis with these two alternative cosmological inference pipelines. We find our results to be consistent with these existing data and analysis pipelines (i.e. the Planck and the DES Y3 weak lensing likelihoods). We recover values lower than Planck not only for $S_8$ \citep[such a tension between CMB and weak-lensing results is already well-known, e.g.][]{Amon_2022} but also for $\Om$.

Table~\ref{tab:results_comparison} presents credible intervals derived from the one-dimensional marginals in Fig.~\ref{fig:results_data_comparison}.

The Planck reanalysis uses the 2018 TTTEEE-lowE likelihood~\cite{aghanim2020planck} with settings matching those used in~\cite{Abbott_2023}. All priors, except for $\Omega_{\rm b}$, were matched to our analysis (Table~\ref{tab:priors}). As our choice of $\Omega_{\rm b}$ prior was informed by Planck, for the Planck reanalysis we use a prior matching~\cite{Abbott_2023}.

We do not include shear ratio information~\citep[e.g.][]{y3-shearratio} for the DES Year 3 weak lensing reanalysis.

We sample the posterior using the Planck and the DES Y3 weak lensing likelihoods with \textsc{PolyChord}~\citep{Handley_2015}. For the parameters $h$, $\Omega_{\rm b}$, $n_s$, and $\Om$, we use a flat prior during the MCMC sampling and then importance re-weight to the desired prior as a post-processing step.

\begin{table}
{ \small
\begin{tabular} { l  c c c}

\hline
\hline

 &  Power spectrum & Peaks \& power  & Map-level DES Y3    \\
  &(this work): $C_\ell$  &(this work): \CLPEAKS{} &    (this work): \CLCNN{}  \\
\hline
\\
$\Om      $ & $0.352^{+0.035}_{-0.053}$ &$0.340^{+0.030}_{-0.042}   $ & $0.283^{+0.020}_{-0.027}$ 
\vspace{0.2cm}\\ 

\vspace{0.2cm}

{$S_8            $} &$0.807^{+0.027}_{-0.025}   $ &$0.807\pm 0.023            $ &  $0.804^{+0.025}_{-0.017}   $ \\

{$w              $} & $<-0.661   $ & $< -0.740                  $&  $< -0.803                  $ \\
\hline
\end{tabular}
}
\caption{\label{tab:results} Comparison of summary statistics used in this analysis (power spectrum, peaks, and map-level inference): 68 per cent credible intervals from the marginal posterior probability distributions of $\Om$, $S_8$, and $w$.}
\end{table}

\begin{table}
{ \small
\begin{tabular} { l  c c c}

\hline
\hline

 &  Map-level DES Y3   & DES Y3 lensing  & Planck (CMB)  \\
  &  (this work): \CLCNN{}  &  likelihood$^*$ & likelihood$^*$ \\
\hline
\\
$\Om      $ & $0.283^{+0.020}_{-0.027}$ & $0.303^{+0.040}_{-0.051}$ & $0.328^{+0.009}_{-0.013} $
\vspace{0.2cm}\\ 

\vspace{0.2cm}

{$S_8            $} & $0.804^{+0.025}_{-0.017}   $ & $0.813^{+0.020}_{-0.029}   $& $0.831^{+0.014}_{-0.015}   $\\

{$w              $} & $< -0.803                  $ & $< -0.707                  $ & $<-0.954 $\\
\hline
& & & $^*$ reanalysed
\end{tabular}
}
\caption{\label{tab:results_comparison} Comparison with existing analyses: 68 per cent credible intervals from the marginal posterior probability distributions of $\Om$, $S_8$, and $w$. We compare the \CLCNN{} result (map-level compression) with the results from both the standard DES weak gravitational lensing (2-point correlation function) likelihood and the Planck CMB data. The standard DES likelihood and Planck likelihood results have used prior choices matched to our analysis to allow comparison.}
\end{table}

\newpage 

\section{Conclusion}\label{sec:conclusion}
We have presented the DES Y3 simulation-based inference results, in which we have used power spectra, peak counts, and map-level compression/inference to constrain parameters of the wCDM model.

Our approach seeks to improve both accuracy and precision.

For improved accuracy we use simulation-based inference as this allows us to forward model realistic effects in our simulated data. For those effects about which there is uncertainty (measurement biases, photometric redshift uncertainties, effects of neutrinos, and intrinsic alignments of galaxies), we randomly vary the effect in our mock data according to our prior probability. This is relatively straightforward in this inference framework; for example, the marginalization over possible redshift distributions $n(z)$ amounts to the marginalization of approximately one thousand nuisance parameters.  

We have tested that our results are robust to certain types of model misspecification (namely source galaxy biasing and baryon feedback). We have also tested that our recovered posterior distributions have the correct coverage; this is made possible by our fast (almost amortized) inference pipeline.

For improved precision we include weak lensing statistics beyond standard two-point statistics. In particular, we directly compress the weak lensing mass map (i.e. \textit{dark matter map}, ~\citealt{KaiserSquires}), and then use simulation-based inference to construct a likelihood for the compressed map. Combining this compressed mass map and the compressed power spectra yields improved constraints on the parameters of the wCDM model (compared to our results using compressed power spectra alone). Table~\ref{tab:results} lists the 68 per cent credible intervals of the marginal posteriors per parameter. 

These improvements are often quoted in terms of the Figure of Merit, given by ${\mathrm{FoM}} = ( {\textrm{det}} \Sigma )^{-1/2}$ for posterior covariance $\Sigma$; this is a measure of inverse volume (i.e. tightness) of the posterior probability. For the weak lensing parameter combination \{$S_8$, $\Om$\} we improve the ${\mathrm{FoM}}$ by a factor of 2.26, while for the dark energy parameter combination \{$\Omega_{\rm DE}$, $w$\} we improve the ${\mathrm{FoM}}$ by a factor of 2.48. (In this latter parameter combination we included neutrinos in $\Om$ and neglected the small photon radiation contribution, so $\Omega_{\rm DE} = 1 - \Om$ for a flat Universe.)

The improvements in precision bring an increased responsibility to maintain accuracy. The principal challenge presented by this approach is achieving sufficiently realistic data modelling, which, if accomplished, will substantially increase the potential for discovery.

\section*{Data Availability}

The Gower Street simulations are available at~\url{www.star.ucl.ac.uk/GowerStreetSims/}; the metacalibration lensing catalogue is available at~\url{https://des.ncsa.illinois.edu}.

The MCMC samples from the parameter posteriors (i.e. chains) will be made available upon publication of the accepted paper.

\section*{Acknowledgements}

We thank F. Lanusse and B. Wandelt for helpful comments and discussions at many points during this project.

 NJ is supported by STFC Consolidated Grant ST/V000780/1 and by the Simons Collaboration on Learning the Universe. The Gower Street simulations were generated under the DiRAC project p153 `Likelihood-free inference with the Dark Energy Survey' (ACSP255/ACSC1) using DiRAC (STFC) HPC facilities (\url{www.dirac.ac.uk}).

 JP has been supported by the Eric and Wendy Schmidt AI in Science Postdoctoral Fellowship, a Schmidt Futures program. JA has been supported by funding from the European Research Council (ERC) under the European Union's Horizon 2020 research and innovation programmes (grant agreement no. 101018897 CosmicExplorer).

 This research used resources of the National Energy Research Scientific Computing Center (NERSC), a U.S. Department of Energy Office of Science User Facility located at Lawrence Berkeley National Laboratory, operated under Contract No. DE-AC02-05CH11231 using NERSC award HEP-ERCAP-0027266.

 Funding for the DES Projects has been provided by the U.S. Department of Energy, the U.S. National Science Foundation, the Ministry of Science and Education of Spain, 
the Science and Technology Facilities Council of the United Kingdom, the Higher Education Funding Council for England, the National Center for Supercomputing 
Applications at the University of Illinois at Urbana-Champaign, the Kavli Institute of Cosmological Physics at the University of Chicago, 
the Center for Cosmology and Astro-Particle Physics at the Ohio State University,
the Mitchell Institute for Fundamental Physics and Astronomy at Texas A\&M University, Financiadora de Estudos e Projetos, 
Funda{\c c}{\~a}o Carlos Chagas Filho de Amparo {\`a} Pesquisa do Estado do Rio de Janeiro, Conselho Nacional de Desenvolvimento Cient{\'i}fico e Tecnol{\'o}gico and 
the Minist{\'e}rio da Ci{\^e}ncia, Tecnologia e Inova{\c c}{\~a}o, the Deutsche Forschungsgemeinschaft, and the Collaborating Institutions in the Dark Energy Survey. 

The Collaborating Institutions are Argonne National Laboratory, the University of California at Santa Cruz, the University of Cambridge, Centro de Investigaciones Energ{\'e}ticas, 
Medioambientales y Tecnol{\'o}gicas-Madrid, the University of Chicago, University College London, the DES-Brazil Consortium, the University of Edinburgh, 
the Eidgen{\"o}ssische Technische Hochschule (ETH) Z{\"u}rich, 
Fermi National Accelerator Laboratory, the University of Illinois at Urbana-Champaign, the Institut de Ci{\`e}ncies de l'Espai (IEEC/CSIC), 
the Institut de F{\'i}sica d'Altes Energies, Lawrence Berkeley National Laboratory, the Ludwig-Maximilians Universit{\"a}t M{\"u}nchen and the associated Excellence Cluster Universe, 
the University of Michigan, NFS's NOIRLab, the University of Nottingham, the Ohio State University, the University of Pennsylvania, the University of Portsmouth, 
SLAC National Accelerator Laboratory, Stanford University, the University of Sussex, Texas A\&M University, and the OzDES Membership Consortium.

Based in part on observations at Cerro Tololo Inter-American Observatory at NSF's NOIRLab (NOIRLab Prop. ID 2012B-0001; PI: J. Frieman), which is managed by the Association of Universities for Research in Astronomy (AURA) under a cooperative agreement with the National Science Foundation.

The DES data management system is supported by the National Science Foundation under Grant Numbers AST-1138766 and AST-1536171.
The DES participants from Spanish institutions are partially supported by MICINN under grants ESP2017-89838, PGC2018-094773, PGC2018-102021, SEV-2016-0588, SEV-2016-0597, and MDM-2015-0509, some of which include ERDF funds from the European Union. IFAE is partially funded by the CERCA program of the Generalitat de Catalunya.
Research leading to these results has received funding from the European Research
Council under the European Union's Seventh Framework Program (FP7/2007-2013) including ERC grant agreements 240672, 291329, and 306478.
We  acknowledge support from the Brazilian Instituto Nacional de Ci\^encia
e Tecnologia (INCT) do e-Universo (CNPq grant 465376/2014-2).

This manuscript has been authored by Fermi Research Alliance, LLC under Contract No. DE-AC02-07CH11359 with the U.S. Department of Energy, Office of Science, Office of High Energy Physics.



\bibliographystyle{mnras_2author}
\bibliography{bibliography,des_y3kp}




\appendix

\section{Author affiliations} \label{append:affiliations}
{
\scriptsize
$^{1}$ Department of Physics \& Astronomy, University College London, Gower Street, London, WC1E 6BT, UK\\
$^{2}$ Department of Physics and Astronomy, University of Pennsylvania, Philadelphia, PA 19104, USA\\
$^{3}$ Oskar Klein Centre for Cosmoparticle Physics, Stockholm University, Stockholm SE-106 91, Sweden\\
$^{4}$ Ruhr University Bochum, Faculty of Physics and Astronomy, Astronomical Institute, German Centre for Cosmological Lensing, 44780 Bochum, Germany\\
$^{5}$  Nordita, KTH Royal Institute of Technology and Stockholm University, Hannes Alfv\'ens v\"ag 12, SE-10691 Stockholm, Sweden\\
$^{6}$ Department of Astronomy and Astrophysics, University of Chicago, Chicago, IL 60637, USA\\
$^{7}$ Kavli Institute for Cosmological Physics, University of Chicago, Chicago, IL 60637, USA\\
$^{8}$ Universit\'e Grenoble Alpes, CNRS, LPSC-IN2P3, 38000 Grenoble, France\\
$^{9}$ Department of Physics, ETH Zurich, Wolfgang-Pauli-Strasse 16, CH-8093 Zurich, Switzerland\\
$^{10}$ Argonne National Laboratory, 9700 South Cass Avenue, Lemont, IL 60439, USA\\
$^{11}$ Institute of Space Sciences (ICE, CSIC),  Campus UAB, Carrer de Can Magrans, s/n,  08193 Barcelona, Spain\\
$^{12}$ Institute of Astronomy, University of Cambridge, Madingley Road, Cambridge CB3 0HA, UK\\
$^{13}$ Kavli Institute for Cosmology, University of Cambridge, Madingley Road, Cambridge CB3 0HA, UK\\
$^{14}$ Physics Department, 2320 Chamberlin Hall, University of Wisconsin-Madison, 1150 University Avenue Madison, WI  53706-1390, USA\\
$^{15}$ Department of Physics, Carnegie Mellon University, Pittsburgh, Pennsylvania 15312, USA\\
$^{16}$ Instituto de Astrofisica de Canarias, E-38205 La Laguna, Tenerife, Spain\\
$^{17}$ Laborat\'orio Interinstitucional de e-Astronomia - LIneA, Rua Gal. Jos\'e Cristino 77, Rio de Janeiro, RJ - 20921-400, Brazil\\
$^{18}$ Universidad de La Laguna, Dpto. Astrofísica, E-38206 La Laguna, Tenerife, Spain\\
$^{19}$ Department of Physics, Duke University Durham, NC 27708, USA\\
$^{20}$ NASA Goddard Space Flight Center, 8800 Greenbelt Rd, Greenbelt, MD 20771, USA\\
$^{21}$ Lawrence Berkeley National Laboratory, 1 Cyclotron Road, Berkeley, CA 94720, USA\\
$^{22}$ Fermi National Accelerator Laboratory, P. O. Box 500, Batavia, IL 60510, USA\\
$^{23}$ Jet Propulsion Laboratory, California Institute of Technology, 4800 Oak Grove Dr., Pasadena, CA 91109, USA\\
$^{24}$ SLAC National Accelerator Laboratory, Menlo Park, CA 94025, USA\\
$^{25}$ University Observatory, Faculty of Physics, Ludwig-Maximilians-Universit\"at, Scheinerstr. 1, 81679 Munich, Germany\\
$^{26}$ Center for Astrophysical Surveys, National Center for Supercomputing Applications, 1205 West Clark St., Urbana, IL 61801, USA\\
$^{27}$ Department of Astronomy, University of Illinois at Urbana-Champaign, 1002 W. Green Street, Urbana, IL 61801, USA\\
$^{28}$ Kavli Institute for Particle Astrophysics \& Cosmology, P. O. Box 2450, Stanford University, Stanford, CA 94305, USA\\
$^{29}$ Department of Astrophysical Sciences, Princeton University, Peyton Hall, Princeton, NJ 08544, USA\\
$^{30}$ Instituto de F\'isica Gleb Wataghin, Universidade Estadual de Campinas, 13083-859, Campinas, SP, Brazil\\
$^{31}$ Department of Physics, University of Genova and INFN, Via Dodecaneso 33, 16146, Genova, Italy\\
$^{32}$ Jodrell Bank Center for Astrophysics, School of Physics and Astronomy, University of Manchester, Oxford Road, Manchester, M13 9PL, UK\\
$^{33}$ Centro de Investigaciones Energ\'eticas, Medioambientales y Tecnol\'ogicas (CIEMAT), Madrid, Spain\\
$^{34}$ Brookhaven National Laboratory, Bldg 510, Upton, NY 11973, USA\\
$^{35}$ Department of Physics and Astronomy, Stony Brook University, Stony Brook, NY 11794, USA\\
$^{36}$ Institut de Recherche en Astrophysique et Plan\'etologie (IRAP), Universit\'e de Toulouse, CNRS, UPS, CNES, 14 Av. Edouard Belin, 31400 Toulouse, France\\
$^{37}$ Excellence Cluster Origins, Boltzmannstr.\ 2, 85748 Garching, Germany\\
$^{38}$ Max Planck Institute for Extraterrestrial Physics, Giessenbachstrasse, 85748 Garching, Germany\\
$^{39}$ Universit\"ats-Sternwarte, Fakult\"at f\"ur Physik, Ludwig-Maximilians Universit\"at M\"unchen, Scheinerstr. 1, 81679 M\"unchen, Germany\\
$^{40}$ Institute for Astronomy, University of Edinburgh, Edinburgh EH9 3HJ, UK\\
$^{41}$ Department of Physics, University of Michigan, Ann Arbor, MI 48109, USA\\
$^{42}$ Institute of Cosmology and Gravitation, University of Portsmouth, Portsmouth, PO1 3FX, UK\\
$^{43}$ School of Mathematics and Physics, University of Queensland,  Brisbane, QLD 4072, Australia\\
$^{44}$ Department of Physics, IIT Hyderabad, Kandi, Telangana 502285, India\\
$^{45}$ Institute of Theoretical Astrophysics, University of Oslo. P.O. Box 1029 Blindern, NO-0315 Oslo, Norway\\
$^{46}$ Instituto de Fisica Teorica UAM/CSIC, Universidad Autonoma de Madrid, 28049 Madrid, Spain\\
$^{47}$ Institut d'Estudis Espacials de Catalunya (IEEC), 08034 Barcelona, Spain\\
$^{48}$ Institut de F\'{\i}sica d'Altes Energies (IFAE), The Barcelona Institute of Science and Technology, Campus UAB, 08193 Bellaterra (Barcelona) Spain\\
$^{49}$ Santa Cruz Institute for Particle Physics, Santa Cruz, CA 95064, USA\\
$^{50}$ Center for Cosmology and Astro-Particle Physics, The Ohio State University, Columbus, OH 43210, USA\\
$^{51}$ Department of Physics, The Ohio State University, Columbus, OH 43210, USA\\
$^{52}$ Center for Astrophysics $\vert$ Harvard \& Smithsonian, 60 Garden Street, Cambridge, MA 02138, USA\\
$^{53}$ George P. and Cynthia Woods Mitchell Institute for Fundamental Physics and Astronomy, and Department of Physics and Astronomy, Texas A\&M University, College Station, TX 77843,  USA\\
$^{54}$ LPSC Grenoble - 53, Avenue des Martyrs 38026 Grenoble, France\\
$^{55}$ Instituci\'o Catalana de Recerca i Estudis Avan\c{c}ats, E-08010 Barcelona, Spain\\
$^{56}$ Observat\'orio Nacional, Rua Gal. Jos\'e Cristino 77, Rio de Janeiro, RJ - 20921-400, Brazil\\
$^{57}$ School of Physics and Astronomy, University of Southampton,  Southampton, SO17 1BJ, UK\\
$^{58}$ Computer Science and Mathematics Division, Oak Ridge National Laboratory, Oak Ridge, TN 37831, USA\\

}



\bsp	
\label{lastpage}
\end{document}